\documentstyle[12pt]{article}
\begin{document}
\def\theequation{\arabic{section}.\arabic{equation}}
\newcommand{\be}{\begin{equation}}
\newcommand{\ee}{\end{equation}}
\begin{titlepage}
\title{PROBING THE GRAVITATIONAL GEON}
\author{F. I. Cooperstock, V. Faraoni and G. P. Perry  \\ \\
{\small \it Department of Physics and Astronomy, University
of Victoria} \\
{\small \it P.O. Box 3055, Victoria, B.C. V8W 3P6 (Canada)}}
\date{}
\maketitle   \thispagestyle{empty}  \vspace*{1truecm}
\begin{abstract}
The Brill--Hartle gravitational geon construct as a spherical shell of
small amplitude, high frequency gravitational waves is reviewed and
critically analyzed.  The Regge--Wheeler formalism is used to
represent gravitational wave perturbations of the spherical background
as a superposition of tensor spherical harmonics and an attempt is
made to build a non--singular solution to meet the requirements of a
gravitational geon. High--frequency waves are seen to be a necessary
condition for the geon and the field equations are decomposed
accordingly. It is shown that this leads to the impossibility of
forming a spherical gravitational geon. The attempted constructs of
gravitational and electromagnetic geons are contrasted. The spherical
shell in the proposed Brill--Hartle geon does not meet the regularity
conditions required for a non--singular source and hence cannot be
regarded as an adequate geon construct. Since it is the high frequency
attribute which is the essential cause of the geon non--viability, it
is argued that a geon with less symmetry is an unlikely prospect. The
broader implications of the result are discussed with particular
reference to the problem of gravitational energy.
\end{abstract}
\begin{center} {\small   To appear in {\em Int. J. Mod. Phys. D} }
\end{center}
\end{titlepage}   \clearpage
\section{Introduction}

Forty years ago, the geon concept was introduced \cite{Wheeler55}:
zero rest mass field concentrations held together for
long periods of time by their gravitational attraction. Such
constructs were motivated by studies of the
motion of bodies in general relativity. More recent interest arises
from the study of the entropy of radiation \cite{Sorkinetal} and from the
analogy between electromagnetic geons and quark stars \cite{Sokolov}.
Electromagnetic, neutrino and mixed type geons were studied
\cite{Wheeler55}, \cite{PowerWheeler}--\cite{Brill66}
and it was suggested that it should be possible to construct a geon from
gravitational waves \cite{ReggeWheeler}. Brill and Hartle
\cite{BrillHartle} (henceforth referred to as BH) attempted the
construction of a gravitational geon model in detail. Later papers
(\cite{Gerlach,CohenWald}~--~see also \cite{ColemanSmarr}) assumed the
correctness of the BH model. In their approach, BH considered a strongly
curved static or quasi--static ``background geometry'' $\gamma_{\mu \nu}$
on top of which a small ripple $h_{\mu \nu}$ resided, satisfying a linear
wave equation. The wave frequency was assumed to be so high as to create a
sufficiently large effective energy density which served as the source
of the background $\gamma_{\mu\nu}$, taken to be
spherically symmetric on a time average. For their analysis, they took the
Regge--Wheeler \cite{ReggeWheeler} (henceforth referred to as RW)
decomposition of $h_{\mu\nu}$ in a
spherical background in
terms of waves characterized by the usual quantum numbers $l$, $m$
related to the angular momentum operators, and by the frequency
$\omega$. They claimed to have found a solution with a flat--space
spherical
interior, a Schwarzschild exterior and a thin shell separation meant to
be created by high frequency gravitational waves. With the mass $M$
identified from the exterior metric, there would follow an unambiguous
realization of the gravitational geon as described above.

To be complete, however, two conditions must be satisfied. Firstly, the
gravitational geon must be a non--singular solution of the Einstein
equations in vacuum. Any singularities present would indicate the presence
of non--gravitational sources $T_{\mu\nu}$ compactified into points,
curves or surfaces,
negating the desired non--singular purely field structure. Secondly, the
consistency of the solution must be demonstrated, namely that the
background $\gamma_{\mu\nu}$ is consistent with the time--averaged
effective density constructed from $h_{\mu\nu}$ as source
in the region of non--vanishing $h_{\mu\nu}$. Regarding the first
condition, it is straightforward to show that the junction conditions
for regularity are not satisfied by the BH solution and hence as it
stands, cannot be taken as singularity--free. With the first condition
violated, there is no basis for proceeding with a consideration of
the second.

One might reasonably argue that while the given structure is
inadequate as it stands, an expansion of the shell region into one of
finite extent would reveal a well--posed geon solution with both
regularity and consistency. Our analysis is sufficiently general to
include this geometry in which the gravitational field decays
sufficiently rapidly at spatial infinity, and to consider also the
possibility of geons ``leaking'' radiation to the exterior.  Odd high
frequency modes in the RW formalism were analyzed in conjunction with
a static and a time--dependent spherically symmetric background metric
$\gamma_{\mu\nu}$. It was found that the Einstein equations do not
allow a solution with the required characteristics and hence a
spherical gravitational geon cannot exist.  While the even mode case
or a case with a more general geometry than spherical was not yet
analyzed, it would be unexpected that such a geon could be found when
the most primitive case is excluded.  Moreover, the key factor which
leads to the non--existence of the spherical geon is not the spatial
symmetry but rather the high frequency.  This fortifies the
expectation that the result is general.

A brief preliminary description of this work was published in
\cite{geonletter}. The present paper provides details of the
calculations and an expanded study of the gravitational geon
problem. It also goes beyond the insufficient generality that was
analyzed in the earlier paper\footnote{We are grateful to to Dr.\ Paul
Anderson for very helpful discussions in this regard.}. In Sec.~2, we
review the basic mathematical formalism for the construction of
gravitational geons. This is used in Sec.~3 to analyze the proposed BH
solution. In Sec.~4, we attempt the construction of a non--singular
solution for a gravitational geon with spherical symmetry and contrast
the results with the electromagnetic case. It is demonstrated that the
Einstein equations do not permit the realization of the gravitational
geon. In Sec.~5, the results of the previous section are discussed and
the viability of the proposed electromagnetic geon is critically
examined. We conclude with a discussion of the potential ramifications
of these results with reference to the problem of gravitational
energy.

\section{Gravitational geons}

We consider the spacetime metric given by \footnote{The metric signature is
$ -+++ $. We use units in which
$G=c=1$. Greek indices run from 0 to 3 and Latin indices run from 1 to 3
(apart from Appendix~B, where they assume the values 0,~2 and 3).
A comma and a semicolon denote, respectively, ordinary and covariant
differentiation with respect to the background metric. The Ricci
tensor is given by $R_{\mu\nu}=
\Gamma^{\sigma}_{\mu\sigma ,\nu}-\Gamma^{\sigma}_{\mu\nu ,\sigma}+
\Gamma^{\sigma}_{\rho\nu}\Gamma^{\rho}_{\mu\sigma}-
\Gamma^{\rho}_{\mu\nu}\Gamma^{\sigma}_{\rho\sigma}$.}\setcounter{equation}{0}
\be  \label{1}
g_{\mu\nu}=\gamma_{\mu\nu}+h_{\mu\nu} \; ,
\ee
where we assume that $g_{\mu\nu}$ is asymptotically flat, that
$\gamma_{\mu\nu}$ is a static, spherically symmetric, asymptotically
flat metric and $h_{\mu\nu}$ are small perturbations ($|h_{\mu\nu}|<<1$)
representing gravitational waves. In a system of Schwarzschild--like
coordinates $\{x^{\alpha}\}=\{ t,r,\theta,\varphi \}$, the background
metric is given by
\be    \label{2}
\left( \gamma_{\mu\nu} \right)=\mbox{diag} \left(
-\mbox{e}^{\nu},\mbox{e}^{\lambda}, r^2, r^2 \sin^2 \theta \right) \; ,
\ee
where
\be       \label{3}
\lambda=\lambda(r) \; , \;\;\;\;\;\;\;\;\;\; \nu=\nu(r)
\ee
and
\be          \label{4}
h_{\mu\nu}=h_{\mu\nu}(t,r,\theta,\varphi ) \; .
\ee
Following BH, we represent the most general
gravitational wave perturbation $h_{\mu\nu} $ of the spherical background
as a superposition of tensor spherical harmonics:
\be
h_{\mu\nu}=\sum_{l=0}^{+\infty} \sum_{m=-l}^{+l}
\int\limits_{0}^{+\infty} d\omega \, h_{\mu\nu}^{(lm\omega )}
(r,\theta ,\varphi ) \, {\mbox e}^{i\omega t} \; +\; \mbox{c.c.}
\ee
This is justified by the
fact that the dynamics of the gravitational waves in the present
context are governed by the {\em linearized} Einstein equations around
the background $\gamma_{\mu\nu}$ and therefore a superposition
principle holds. Due to linearity, we can restrict ourselves to a study
of the evolution of the single tensor spherical modes. For ease
of comparison with the BH paper, we will
use the RW set of tensor spherical harmonics
(\cite{ReggeWheeler}, \cite{Manasse}--\cite{Sandberg}; see \cite{Thorne}
for a
review and for relations with other sets of tensor spherical
harmonics). An ``even mode'' (also called ``polar mode'' by other authors
\cite{Chandrasekhar})
in the RW formalism is factorized as the product of
functions dependent only on time, radius, and angles respectively. The
angular part is determined by the numbers $l$ and $m$ related to the
usual scalar spherical harmonics. The even modes have the form
\be  \label{4ter}
h_{\mu\nu}^{( \mbox{even})}\left( t,r,\theta,\varphi\right)=\left(
\begin{array}{ccccc}
-\mbox{e}^{\nu} H_0(r) & H_1(r) & 0 &  0 & \\
                &        &   &    & \\
H_1(r) & \mbox{e}^{\lambda}H_2(r)  & 0 &  0 & \\
                &        &   &    & \\
0 & 0 & r^2 K(r) & 0 & \\
                &        &   &    & \\
0 & 0 & 0 & r^2 K(r) \sin^2 \theta & \\
                &        &   &    & \\
\end{array} \right)  Y^{lm} e^{-i\omega t} \; ,
\ee
where $ Y^{lm}( \theta, \varphi)$ are the usual spherical harmonics
\footnote{Strictly speaking, the radial functions in
Eqs.~(\ref{4ter}) and (\ref{4bis}) depend on $\omega$, $l$ and $m$ and
should be labelled accordingly. However, this would result in a
cumbersome notation that is preferably avoided.}.
These modes have parity $(-)^l$. The ``odd modes'' (in the RW
terminology~--~also called ``axial modes'') are given by
\be  \label{4bis}
h_{\mu\nu}^{( \mbox{odd})}\left( t,r,\theta,\varphi\right)=
\left(
\begin{array}{ccccc}
0 & 0 & -h_0(r) \left( \sin \theta \right)^{-1} \, \frac{\partial}
{\partial \varphi} Y^{lm}& h_0(r) \sin \theta
\,\frac{\partial}{\partial \theta}Y^{lm}  & \\ 
                &        &   &    & \\
0 & 0 & -h_1(r) \left( \sin \theta \right)^{-1} \,
\frac{\partial}{\partial \varphi}Y^{lm} 
& h_1(r) \sin \theta \,\frac{\partial}{\partial \theta} Y^{lm}& \\ 
                &        &   &    & \\
\mbox{Sym} & \mbox{Sym} & 0 & 0 & \\
                &        &   &    & \\
\mbox{Sym} & \mbox{Sym} & 0 &  0 & \\
\end{array} \right)  e^{- i\omega t}
\ee
and have parity $(-)^{l+1}$. We will consider the case of odd 
modes only because of the relative ease in computations. It is
unlikely that the even modes would produce a contrary result although
it would be useful if a follow--up calculation were to be performed to
verify this conjecture.

A {\bf gravitational geon} is defined as a bounded configuration of
gravitational waves whose gravity is sufficiently strong to keep them
confined on a time scale long compared to the characteristic composing
wave period. It is required that no matter or fields other
than the gravitational field be present. Although one may consider the
possibility of strong gravitational waves, and the definition of
gravitational geon allows for this possibility, in this paper we will
restrict ourselves to the case in which the amplitude of gravitational
waves is small. This permits us to apply the linearized Einstein
theory to the propagation of each single wave
in the background created by the average action of all the
waves composing the geon. Furthermore, it is required
that the configuration represented by the metric $\gamma_{\mu\nu}$ be
stable over a time scale much larger than the typical period of its
gravitational wave constituents, and that the gravitational field becomes
asymptotically flat at spatial infinity. Gravitational geons
were introduced on
the basis of the analogy with electromagnetic and neutrino geons in
the RW paper and were studied in greater detail by BH.
Wheeler's method of building an electromagnetic geon was to
replace the details of the electromagnetic field by the time average of
the components of the electromagnetic stress--energy tensor. Upon
averaging over many modes of oscillation of the electromagnetic field,
one obtains a stress-energy tensor, and as a consequence, a
gravitational field and metric which are spherically symmetric. Any
given mode of oscillation is taken to propagate in the spherically
symmetric gravitational field created by the rest of the radiation.
The attempt to build a geon resembles the construction, in other
fields of physics, of a system with many (almost) identical components,
each of which introduces a negligible perturbation in the dynamics
of the whole system and has an evolution governed by the averaged action
of all the other components. An
example of such a system in Newtonian theory is a galaxy described by
the potential created by the mass distribution of many stars (here we
neglect dark matter, and the fact that a potential--density pair
usually describes only a single component of a galaxy, and is adequate
only for certain types of galaxies \cite{BinneyTremaine}). Each star
gives a very small contribution to this potential and its orbit is
determined by the global galactic potential.

Consistent with this idea, it is required that
\be    \label{5}
\gamma_{\mu\nu}=\left\langle  g_{\mu\nu} \right\rangle  \; .
\ee
We also have
\be  \label{6}
\left\langle h_{\mu\nu} \right\rangle=\left\langle \frac{\partial
h_{\mu\nu}}{\partial
x^{\alpha}} \right\rangle= \left\langle \frac{\partial^2 h_{\mu\nu}}
{\partial x^{\alpha}\partial x^{\beta}} \right\rangle= 0 \; ,
\ee
where $ \left\langle \right.$~~$\left. \right\rangle $ denotes an average
over a time that is much
longer than the typical gravitational wave wavelength $\lambda$
(``Brill--Hartle average''). A mathematically rigourous treatment of this
concept is contained in the paper by MacCallum and Taub \cite{MacCallumTaub}.
This idea has proved very valuable and the averaging process has been used by
many authors after BH, and is well defined only if it
is assumed that the typical wavelength $\lambda$ \footnote{The term
``typical gravitational wavelength''
$\lambda$ may be source of confusion to some readers. Since
we are decomposing
the general wave form into an infinite set of Regge--Wheeler modes, one may
think that $\lambda$ represents the wavelength of each mode, and that
Eq.~(\ref{7}) is only valid if the geon was composed of one
and only one mode.
However, when one is analyzing a general wave form, it is justifiable
to assign
a {\em single} parameter describing the scale of variation of
the wave form. In
the present context, $\lambda$ is the scale over which the wave form varies.
Equation~(\ref{7}) is easily derived from Eq.~(\ref{239c}) if one keeps
in mind that $h_{\mu\nu ,\alpha}\sim \epsilon /\lambda$ etc. (see
\cite{MTW}) and that
$\lambda$ represents the scale of variation of $h_{\mu\nu}$.}
is much smaller than
the space and time scale of variation $L$ of the background metric
$\gamma_{\mu\nu}$ (high frequency approximation) \cite{Isaacson}:
\be    \label{7}
\epsilon \equiv \frac{\lambda}{L}<< 1 \; .
\ee
This assumption provides us with a smallness parameter $\epsilon$ to
be used as an expansion parameter. Following \cite{Isaacson}, we
measure times and lengths in units of $L$ so that $\lambda=\epsilon$.
We have also
\be   \label{8}
h_{\mu\nu}=\mbox{O}( \epsilon)      \; ,                  \ee
\be   \label{9}
\omega=\frac{2 \pi}{\lambda}=\mbox{O} \left( \, \frac{1}{\epsilon}\,
\right)  \; ,                                           \ee
\be   \label{10}
\mbox{O}\left( \frac{\partial h_{\mu\nu}}{\partial x^{\alpha}}
\right)=\mbox{O}(1) \; ,
\ee
\be         \label{10bis}
\mbox{O}\left( \frac{\partial^2 h_{\mu\nu}}{\partial
x^{\alpha}\partial x^{\beta}} \right)=\mbox{O}\left( \,
\frac{1}{\epsilon}\, \right)    \; .
\ee
In our notation, O(1)$\equiv$O($\epsilon^0$). Equation~(\ref{8}) is derived
in \cite{Isaacson,MTW,LandauLifschitz}. It is to be noted that, in
the most general case of high frequency gravitational waves on a
curved spacetime, two smallness parameters are involved: the
dimensionless amplitude of the waves and the ratio $\lambda/L$. These
two parameters coincide in the specific case under consideration, in which
the only source of the background curvature are the gravitational
waves. One can conceive of situations in which more than one parameter arises
from the high frequency approximation, and these cases have been considered in
the literature (see e.g. \cite{Araujo}). However, in these situations,
gravitational waves are not the only source of curvature. When gravitational
waves are the only source of curvature, as in the gravitational geon, these
multiple parameters reduce to the single parameter $\epsilon$.
Equation~(\ref{10}) implies that the quantum numbers $l$ and $m$
are of order O($1/\epsilon$).

The Ricci tensor can be expanded in the form
\cite{BrillHartle,Isaacson}
\be     \label{11}
R_{\alpha\beta}( g)=R^{(0)}_{\alpha\beta}\, ( \gamma
)+R^{(1)}_{\alpha\beta}\, ( \gamma, h) +R^{(2)}_{\alpha\beta}\,
( \gamma,h) + \cdots  \; ,                               \ee
where (\cite{BrillHartle,Isaacson} and references therein)
\be
R^{(1)}_{\alpha\beta}=\frac{1}{2}\, \gamma^{\rho\tau}\left(
h_{\rho\tau;\alpha\beta}+h_{\alpha\beta;\rho\tau}-
h_{\tau\alpha;\beta\rho}-h_{\tau\beta;\alpha\rho}\right)
\; ,\label{40bis}
\ee
\begin{eqnarray}
& & R^{(2)}_{\alpha\beta}=-\,\frac{1}{2}\, \left[
\frac{{h^{\rho\tau}}_{;\beta}}{2}\, h_{\rho\tau ;\alpha}
+h^{\rho\tau} \left( h_{\tau\rho ;\alpha\beta}
+h_{\alpha\beta ;\rho\tau}-
h_{\tau\alpha ;\beta\rho}-h_{\tau\beta ;\alpha\rho} \right)
\right. \nonumber \\
& & \mbox{}\left.+
{h_{\beta}}^{\tau ;\rho}\left( h_{\tau\alpha ;\rho}-
h_{\rho\alpha ;\tau}\right)
-\left( {h^{\rho\tau}}_{;\rho}-\frac{h^{;\tau}}{2} \right) \left(
h_{\tau\alpha ;\beta}+h_{\tau\beta ;\alpha}-h_{\alpha\beta
;\tau}\right) \right]   \; ,      \label{R2}
\end{eqnarray}
and $h\equiv {h^{\alpha}}_{\alpha}$. The term
$R^{(0)}_{\alpha\beta}\,( \gamma)$ is the Ricci tensor
of the background metric
$\gamma_{\mu\nu}$, whereas $R^{(1)}_{\alpha\beta}$  and
$ R^{(2)}_{\alpha\beta}$ are, respectively, the parts of the Ricci
tensor linear and quadratic in $ h_{\mu\nu}$ and their derivatives. In
the absence of high frequency waves (or on a flat background), $h_{\mu\nu}$
and their derivatives
are all of order O($\epsilon$). In this case the superscripts on the
expansion terms of Eq.~(\ref{11}) also indicate its order in powers of
$\epsilon$. However, in the high frequency approximation it is clear
that $R^{(1)}_{\mu\nu}$ contains terms of order O($1/\epsilon$) and
O(1) as well as O($\epsilon$) \cite{Isaacson}. Similarly,
$R^{(2)}_{\mu\nu}$ is comprised of terms of order O(1), O($\epsilon$),
etc. Solving the vacuum field equations
\be
\label{vacuumefe}
R_{\mu\nu}\left( g\right)=0
\ee
consistently to any order of approximation requires that we set
each order in the expansion parameter $\epsilon$ equal to zero.
We express Eqs.~(\ref{40bis}) and (\ref{R2}) as
\begin{eqnarray}
& & R^{(1)}_{\mu\nu}\left( \gamma, h \right)=
R^{(1)}_{\mu\nu}\left[ \epsilon^{-1} \right] +R^{(1)}_{\mu\nu}\left[
\epsilon^{0} \right] + \cdots  \; ,   \\
& & R^{(2)}_{\mu\nu}\left( \gamma, h \right)=
R^{(2)}_{\mu\nu}\left[ \epsilon^{0} \right] +R^{(2)}_{\mu\nu}\left[
\epsilon \right] + \cdots  \; ,
\end{eqnarray}
where $R^{(k)}_{\mu\nu}\left[ \epsilon^n \right] $ denotes the
term of order O($\epsilon^n$) in $R^{(k)}_{\mu\nu}$. The first order
approximation is thus
\be               \label{239b}
R^{(1)}_{\mu\nu}\left[ \epsilon^{-1} \right]=0 \; .     \ee
The second order approximation requires that terms of order O(1) be
set equal to zero. The field equations to this order are
\be               \label{239a}
R^{(0)}_{\mu\nu}\left( \gamma \right)+R^{(1)}_{\mu\nu}\left[ \epsilon^0
\right]+R^{(2)}_{\mu\nu}\left[ \epsilon^0 \right]=0 \; .
\ee
Performing the Brill--Hartle average on Eq.~(\ref{239a}), one obtains
\be     \label{239c}
R^{(0)}_{\mu\nu}\left( \gamma \right)=-\left\langle R^{(2)}_{\mu\nu}
\left[ \epsilon^0 \right] \right\rangle \; .                \ee
Note that from Eq.~(\ref{6})
\be
\left\langle R^{(1)}_{\mu\nu}\left[ \epsilon^{-1} \right]
\right\rangle=\left\langle R^{(1)}_{\mu\nu}\left[ \epsilon^0 \right]
\right\rangle=\cdots =0              \ee
and hence
\be
\left\langle R^{(1)}_{\mu\nu}\left( \gamma, h \right)
\right\rangle=0 \; .                     \ee
In Eq.~(\ref{239c}) the part of the Ricci tensor quadratic in
$h_{\mu\nu}$ and their derivatives has been taken to the right hand side
and is seen as an effective source term due to the gravitational waves.
It is important
to note that Eq.~(\ref{239c}) has the potential to lead to the
description of a gravitational geon only by virtue of the high
frequency approximation. Under the assumption that
gravitational waves are weak but not of high frequency,
Eqs.~(\ref{9})--(\ref{10bis}) would not hold and the two terms in
Eq.~(\ref{239c}) would have different orders.
$R^{(2)}_{\alpha\beta}=\mbox{O}( \epsilon^2)$ could never balance
$R^{(0)}_{\alpha\beta}( \gamma)=\mbox{O}(1)$ in
this equation. This would prevent {\em a~priori} the construction of a
gravitational geon. This point can be understood physically by noting
that the effective energy density associated with gravitational waves with
amplitude $h<<1$ and frequency $\omega$ is roughly proportional to
$\left( h\omega \right)^2$. This quantity can be of order unity only if
$\omega \sim 1/h >>1$. Therefore, it is clear that the high frequency
approximation is a necessary condition for geon construction in the
present context.

We shall designate as the ``geon problem'', the problem of finding a
solution $(
\gamma_{\mu\nu}, h_{\mu\nu})$ to the Einstein equations (\ref{239b}),
(\ref{239a}) and (\ref{239c}) with the above mentioned properties and
satisfying the boundary conditions describing asymptotic
flatness
\be   \label{17}
h_{\mu\nu} \rightarrow 0 \;\;\;\;\;\;\; \mbox{as}\;\;\;\;\;\;\;
 r \rightarrow +\infty \; .                          \ee

\section{The BH analysis}

To the authors' knowledge the only explicit attempt at gravitational
geon construction was that of BH. In this Section we review
their pioneering approach to the problem and critically analyze their work.

We follow BH in expressing the gravitational wave perturbations in
terms of RW tensor 
spherical harmonics. For the sake of simplicity, as
done by BH, we restrict ourselves to the case of odd modes with zero
angular momentum along the $z$--axis (i.e. $m=0$). The last assumption
eliminates the $\varphi$--dependence from the $h_{\mu\nu}$ functions
and considerably simplifies the Einstein equations. This can be seen
from Eq.~(\ref{4bis}) and from the well--known form of the spherical
harmonics that we present in Eqs.~(\ref{21}), (\ref{21bis}) below.
Thus, the metric perturbations are
\footnote{For ease of comparison with the BH paper, we use
a complex exponential to describe the time--dependence of the metric
perturbations in Eq.~(\ref{18}). This notation is adequate as long as
linear quantities in $h_{\mu\nu}$ and their derivatives are
considered, but clearly it is incorrect when the part of the Ricci tensor
quadratic in $ h_{\mu\nu}$ and their derivatives enters the
discussion. For future reference, we use a function of $\cos \left( \omega
t \right) $ and $\sin \left( \omega t \right) $ instead of a complex
exponential in our calculations of Sec.~4.
}\setcounter{equation}{0}
\be  \label{18}
h_{\mu\nu}\left( t,r,\theta \right)={\cal R}_{\mu\nu}(r) \,
\Theta^l( \theta) \, \mbox{e}^{-i\omega t}     \; ,        \ee
where
\begin{eqnarray}
\label{19}
&& {\cal R}_{\mu\nu}(r) = h_0(r) \left(
\delta^0_{\mu}\,\delta^3_{\nu}+\delta^3_{\mu}\,\delta^0_{\nu}
\right) +
h_1(r) \left( \delta^1_{\mu}\,\delta^3_{\nu}+\delta^3_{\mu}\,
\delta^1_{\nu} \right)         \; ,        \\
\label{20}
&& \Theta^l( \theta) = \sin \theta \,\,
\frac{dY^{l0}}{d\theta}= 
C^{l0} \,\sin \theta \,\, \frac{d P^l ( \cos \theta)}{d \theta}
\; .                     \end{eqnarray}
Here we use the expression of the spherical harmonics
\be    \label{21}
Y^{lm}( \theta, \varphi)=C^{lm} \mbox{e}^{im\varphi} P^{lm}( \cos \theta)
\;\;\;\;\;\;\;\;\;\;(m \geq 0 )    \; ,                        \ee
\be    \label{21bis}
Y^{lm}( \theta, \varphi)=(-1)^m \left( Y^{l|m|} \right)^*
\;\;\;\;\;\;\;\;\;\;(m<0 )   \; ,                         \ee
where $C^{lm}$ are normalization constants. Here ${}^*$ denotes complex
conjugation and $P^{lm}(x)$ are the associated
Legendre polynomials (which can be expressed in terms of the Legendre
polynomials $P^l(x)$). Using the relation $P^{l0}(x)=P^l(x)$ we obtain
\be   \label{23}
Y^{l0}( \theta)=C^{l0} P^l ( \cos \theta)
\; ,                                \ee
from which Eq.~(\ref{20}) follows \footnote{Note a misprint in the second
of the equations~(8)
in \cite{BrillHartle}, corresponding to our Eq.~(\ref{19}). Also to be noted
is an inconsistency in the notation therein: the form
(\ref{18})--(\ref{20}) for the metric perturbations is assumed in
\cite{BrillHartle},
but the
number $m$ in the definition of the function $\Theta^{lm}$ corresponding
to our $\Theta^l$ is retained. This is inappropriate since it
is clear from Eqs.~(8) and (9) in \cite{BrillHartle} that the
intention was to set
$m=0$. Otherwise, the function $\Theta^{lm}$ would depend
on both $\theta$ and $\varphi$, which
is not the case, and the Einstein equations would be much more complicated.
}.

One can now insert the form~(\ref{18})--(\ref{20}) of the metric
perturbations into the Einstein equations (\ref{vacuumefe}), obtaining
equations for the unknown functions $h_0(r)$ and $h_1(r)$. Simultaneously
solving Eqs.~(\ref{239b}) and (\ref{239c}) for a pair $(
\gamma_{\mu\nu}, h_{\mu\nu})$ then provides a solution to the geon
problem.

The correct order of magnitude of the various terms in the
Einstein equations is determined by Eqs.~(\ref{8})--(\ref{10bis}).
The correct order of magnitude decomposition of the
Einstein equations is absent in \cite{BrillHartle}. While the high frequency
approximation was assumed in \cite{BrillHartle},
it was not incorporated into the calculations. As a result, the authors
did not obtain the two
different orders O$( 1/\epsilon )$ and O(1) in the Einstein
equations, using a parameter
$\epsilon$ arising from the high frequency approximation. This is evident
from the fact that their final equations~(10a)--(10c) and (14) contain
terms of different orders in the high frequency limit. In
the remaining part of this Section we will show how the BH results can
be reproduced and we will comment on their proposed geon model.

The BH equations can only be reproduced in the absence of high
frequency waves. In terms of
a parameter $\epsilon$ related to the weakness of the gravitational
waves, Eqs.~(\ref{8})--(\ref{10bis}) must be replaced by
\be  \label{25}
\mbox{O}\left( h_{\mu\nu}\right)=\mbox{O}\left( \frac{\partial h_{\mu\nu}}
{\partial x^{\alpha}}\right)=
\mbox{O}\left( \frac{\partial^2 h_{\mu\nu}}{\partial
x^{\alpha}\partial x^{\beta}} \right)=\mbox{O}( \epsilon)
\;\;\;\;\;\;\; \alpha\; , \;\beta=0,...,3 \; .
\ee
As a consequence of these equations, the Ricci tensor has the form
given by Eq.~(\ref{11}), where $R^{(0)}_{\mu\nu}( \gamma)=$O(1),
$R^{(1)}_{\mu\nu}=\mbox{O}( \epsilon)$ and
$R^{(2)}_{\mu\nu}=\mbox{O}\left( \epsilon^2 \right)$. To order O(1)
the Einstein equations give the well--known
equations for a spherically symmetric, static background (see e.g.
\cite{LandauLifschitz}, p.~300) with vanishing energy--momentum
tensor. As far as the order O$( \epsilon)$ is concerned, only the
(0,~3), (1,~3) and (2,~3) components of the Ricci tensor give nontrivial
results. These components are
\begin{eqnarray}
&& R^{(1)}_{03}=-\,\frac{\mbox{e}^{-\lambda}}{2} \left[ \dot{h}_{13}
\left( \frac{2}{r}-\frac{\lambda'}{2}-\frac{\nu'}{2} \right)
+\frac{h_{03}'}{2} \left(  \lambda'+\nu' \right)
+\dot{h}_{13}'-h_{03}''-\frac{2\nu'}{r}\, h_{03} \right]  \nonumber \\
&& +\,\frac{1}{2r^2} \left( h_{03,22}-h_{03,2} \cot \theta \right)
\; ,  \label{28}                                    \\
\label{29}
&& R^{(1)}_{13}=-\,\frac{\mbox{e}^{-\nu}}{2} \left( \ddot{h}_{13}-
\dot{h}_{03}'+\frac{2\dot{h}_{03}}{r}\right)+
\frac{\mbox{e}^{-\lambda}}{r}\, h_{13}
\left( \frac{\lambda'}{2}-\frac{\nu'}{2}-\frac{1}{r} \right) \nonumber
\\
&& +\frac{1}{2r^2} \left( h_{13,22}-h_{13,2} \cot \theta \right) \; , \\
&& R^{(1)}_{23}=-\,\frac{\mbox{e}^{-\lambda}}{2} \left[
h_{13,2}'-2h_{13}' \cot \theta +
h_{13} \left( \lambda'-\nu'\right) \cot \theta +
\frac{h_{13,2}}{2} \left( \nu'-\lambda' \right) \right]   \nonumber \\
&& -\mbox{e}^{-\nu} \left( \dot{h}_{03} \cot \theta -
\frac{\dot{h}_{03,2}}{2}\right)\; ,  \label{30}
\end{eqnarray}
where a dot and a prime denote differentiation with respect to $t$ and
$r$, respectively. We now insert the form of the metric perturbations
(\ref{18})--(\ref{20}) into the Einstein equations~(\ref{239b}) and
use the following property of the function $\Theta^l$ (see Appendix~A):
\be   \label{31}
\frac{d^2 \Theta^l}{d \theta^2}-\cot \theta \, \frac{d \Theta^l}{d
\theta}+l(l+1) \, \Theta^l=0   \; .            \ee
After some manipulations we find
\footnote{See 
Ref.~\cite{EdelsteinVishveshwara} for corrections to the radial equations in 
Refs.~\cite{ReggeWheeler,BrillHartle}.
Also note misprints in the BH
Eq.~(11) corresponding to our Eq.~(\ref{35}). One of the coefficients
of $Q$ in our Eq.~(\ref{39}) differs by a factor $1/2$ from the
corresponding one in BH Eq.~(14). The sign of the right hand side of
our Eq.~(\ref{38}) is opposite to that in the corresponding BH equation.}
\be   \label{32}
i\omega \left[ h_1'+h_1\left(
\frac{2}{r}-\frac{\lambda'}{2}-\frac{\nu'}{2} \right)
\right]-\frac{h_0'}{2} \left( \lambda'+\nu'\right)+h_0''-h_0\left[
l(l+1) \, \frac{\mbox{e}^{\lambda}}{r^2}-\frac{2\nu'}{r} \right]=0  \; ,
\ee
\be   \label{33}
i\omega \, \mbox{e}^{-\nu}\left( h_0'-\frac{2h_0}{r}\right)+h_1\left[
\frac{l(l+1)}{r^2}-\omega^2\mbox{e}^{-\nu}
+\frac{\mbox{e}^{-\lambda}}{r}\left(
\lambda'-\nu'-\frac{2}{r}\right) \right]=0     \; ,        \ee
\be    \label{34}
i\omega \,\mbox{e}^{-\nu}h_0+\mbox{e}^{-\lambda}\left[ h_1'
+\frac{h_1}{2}\left(
\nu'-\lambda' \right) \right]=0    \; .                         \ee
Following BH we can now use Eq.~(\ref{34}) to eliminate $h_0$ from
Eq.~(\ref{33}), obtaining the second order differential equation for
$h_1(r)$:
\be       \label{35}
h_1''+h_1' \left[ \frac{3}{2} \left( \nu'-\lambda' \right)-\frac{2}{r}
\right] +h_1 \left[ \frac{1}{2} \left(
\nu'-\lambda'\right)^2+\frac{1}{2} \left( \nu''-\lambda''
\right)-l(l+1) \, \frac{\mbox{e}^{\lambda}}{r^2} +\omega^2
\mbox{e}^{\lambda-\nu}+\frac{2}{r^2} \right]=0    \; .
\ee
We introduce the variable $Q$ and the Regge--Wheeler coordinate
$r_*$ defined by
\be \label{36}
h_1 \equiv r \mbox{e}^{( \lambda-\nu )/2}\, Q  \; ,         \ee
\be \label{37}
dr_*=\mbox{e}^{( \lambda-\nu) /2} \, dr   \; .              \ee
In terms of these quantities we have
\be   \label{38}
h_0=-\, \frac{1}{i\omega}\, \frac{d( rQ)}{dr_*}
\ee
and\footnote{An equation similar to Eq.~(\ref{39}) can be
derived for the even modes with $m=0$ \cite{Zerilli2}.}

\be \label{39}
\frac{d^2Q}{dr_*^2}+\left[ \omega^2+\frac{3}{2r}\left( \nu'-\lambda'
\right) \mbox{e}^{\nu-\lambda}-\frac{l(l+1)}{r^2} \,
\mbox{e}^{\nu} \right]
Q=0 \; .                                                   \ee
This Schr\"odinger--like equation lends itself to the analogy with the
dynamics of waves propagating in an effective potential
\cite{Wheeler55,ReggeWheeler,BrillHartle}.

At this point BH proceed with the specification of the
background metric
\be
\mbox{e}^{\nu} =\left\{ \begin{array}{cllll}
1/9 & & \;\;\; \mbox{if}  & r\leq a & \nonumber \\
    & &                   &          & \nonumber \\
1-2M/r & &\;\;\; \mbox{if}\,\,\,\, & r\geq a & \nonumber \\
\end{array} \right.           \; ,                     \label{star1}
\ee
\be
\mbox{e}^{\lambda} =\left\{ \begin{array}{cllll}
1 & & \;\;\;\;\; \mbox{if}  & r< a & \nonumber \\
    & &                   &          & \nonumber \\
\left( 1-2M/r\right)^{-1} & &\;\;\;\;\; \mbox{if}\,\,\,\, & r> a &
\nonumber \\
\end{array} \right.          \; ,                      \label{star2}
\ee
where $ a=9M/4$ and $M$ is the geon mass. This vacuum solution for the
background metric implies that the effective energy density due to the
gravitational waves vanishes for $r\neq a$. Since the effective energy
is positive semi--definite, Eqs.~(\ref{star1}), (\ref{star2}) imply that
\be             \label{1000}
h_{\mu\nu}=0 \;\;\;\;\;\; \mbox{for} \;\;\;\; r\neq a \; .
\ee
Conversely, if the condition (\ref{1000}) is satisfied, the Birkhoff
theorem guarantees that the metric is Minkowskian for $r<a$ and the
Schwarzschild metric for $r>a$.

Therefore, in the BH model, gravitational waves are confined
to a spherical shell, the thickness of which is exactly zero.
Apparently, BH meant to build a geon model in which the
gravitational waves are trapped in a spherical shell which has a
nonvanishing thickness which is much smaller than its radius.
However, their equations do not allow for this possibility. To be
complete, we examine the viability of a geon with gravitational waves
confined
to a shell whose thickness is
exactly zero. It is easy to see that such a model is physically
meaningless and that the geon problem becomes mathematically
ill--defined in this case. In fact, the solutions of the radial
equations (\ref{32})--(\ref{35}) cannot be ordinary functions but must
be sought in some space of {\em distributions}.  In Eq.~(\ref{35}),
the coefficients proportional to $\nu'-\lambda'$ and $\nu''-\lambda''$
are not ordinary functions and have a mathematical meaning only if
they are regarded as distributions. The first of these two quantities
can be expressed as
\be
\nu'-\lambda'=4M r^{-2} \left( 1-\frac{2M}{r}\right)^{-1} \theta_H
\left( r-a \right)   \; ,                                \ee
where
\be
\theta_H (x) \equiv \left\{ \begin{array}{cllll}
0 & & \;\;\; \mbox{if}  & x<0 & \nonumber \\
  & &                   &     & \nonumber \\
1 & &\;\;\; \mbox{if}\,\,\,\, & x>0 & \nonumber \\
\end{array} \right.                          \label{Heaviside}
\ee
is the Heaviside step function. Clearly, the radial derivative of $
\nu'-\lambda'$ can be taken only in a distributional sense. Therefore
the solutions of the Einstein equations are distributions and their
domain is some space of test functions which must be specified in such
a way that the coefficients and the operations involved in the
Einstein equations are well defined. There is no indication as to the
manner in which this functional space should be determined. It seems
almost certain that, if a
meaningful and unambiguous mathematical formulation of the problem
can be
given, the distributional solutions $h_{\mu\nu}$ cannot be seen as
locally integrable functions, but rather must have properties like
a Dirac delta with support on $r=a$. Furthermore, the product of
distributions is not defined and the Einstein equations involving the
part of the Ricci tensor quadratic in $h_{\mu\nu}$ and its derivatives
is mathematically meaningless in this case. This destroys the
possibility of exploring one of the essential features of a
gravitational geon. Moreover, if the $h_{\mu\nu}$ are allowed to be
distributions, the whole meaning of the linearization around the
background $\gamma_{\mu\nu}$, the condition $|h_{\mu\nu}|<<1$, and the
estimates of the different orders of magnitude in the Einstein
equations, become
meaningless. The physical interpretation of a distributional metric
and Riemann tensor is problematic. To appreciate this, one can
consider the much simpler case of a metric which does not satisfy
the appropriate junction conditions \cite{junction} on a spacelike or
timelike hypersurface (this is the case of the metric
$\gamma_{\mu\nu}$ given by Eqs.~(\ref{star1}), (\ref{star2}) and the
timelike hypersurface $r=a$~--~see Appendix~B). As suggested by Israel
\cite{Israel},
and as can be seen from the computation of the Einstein tensor for the
spherical metric specified by Eqs.~(\ref{star1}), (\ref{star2}), a
singular hypersurface $S$ (in the sense \cite{junction} that the first,
or the second fundamental form, or both are not continuous at $S$)
is associated with nonvanishing $T_{\mu\nu}$, a source of stresses.
The definition of a geon, a structure of pure
gravitational waves in the absence of matter, excludes the use of a
background metric which does not satisfy the proper junction
conditions. If, in addition, the ``perturbations'' $h_{\mu\nu}$ are
allowed to be distributions, the consideration of junction conditions
loses its meaning, but the argument shows that delta--like
sources of stresses are included in the problem. Thus, we
exclude the case in which gravitational waves are confined to a shell,
the thickness of which is exactly zero, as physically meaningless,
mathematically ill--defined, and nonviable.

The only possible alternative for a geon model in which gravitational
waves are confined to a spherical shell is the case in which the shell
has a nonzero thickness. Apparently, BH meant to consider
such a model, although this contradicts some of their
equations. To be specific, let us consider a shell of radius $a$ and
thickness $\delta a $ described by values of the radial coordinate in the
range
\be
a-\frac{\delta a}{2} \leq r \leq  a+\frac{\delta a}{2} \; ,
\ee
where $0<\delta a <<a$. In order for the geon to be a distribution of
pure gravitational fields without matter, we must require that the
metric tensor satisfies the appropriate junction conditions
\cite{junction} at the two timelike hypersurfaces
$S_{\pm}=\left\{ \left(t,r,\theta,\varphi \right) :
\;\; r=a\pm \delta a/2 \right\}$. This
guarantees the absence of a real (as opposed to ``effective'', i.e.
generated by gravitational waves) stress--energy tensor $T_{\mu\nu}$
representing a matter distribution. In this model, the modified BH
solution would be
\be
\mbox{e}^{\nu} =\left\{ \begin{array}{cllll}
1/9 & & \;\;\; \mbox{if}  & r\leq a-\delta a/2 & \nonumber \\
    & &                   &                    & \nonumber \\
1-2M/r & &\;\;\; \mbox{if}\,\,\,\, & r\geq a+\delta a/2
& \nonumber \\
\end{array} \right.           \; ,             \label{newstar1}
\ee
\be
\mbox{e}^{\lambda} =\left\{ \begin{array}{cllll}
1 & & \;\;\; \mbox{if}  & r\leq a-\delta a/2 & \nonumber \\
  & &                   &                    & \nonumber \\
\left( 1-2M/r\right)^{-1} & &\;\;\; \mbox{if}\,\,\,\, & r\geq a
+\delta
a/2 & \nonumber \\
\end{array} \right.               \; ,           \label{newstar2}
\ee
\be
h_{\mu\nu}=0 \;\;\;\;\; \mbox{if} \;\;\;\;\; r<a-\frac{\delta a}{2},
\;\;\;\;\;r>a+\frac{\delta a}{2} \; .
\ee
The form of the background metric $\gamma_{\mu\nu}$ inside the
spherical shell is not given by BH and must be determined by solving
simultaneously the Einstein equations to the two lowest orders for a
pair $\left( \gamma_{\mu\nu}, h_{\mu\nu} \right)$ \cite{Araujo}. The
proper orders of magnitude did not appear in \cite{BrillHartle} as a
consequence of neglecting the high frequency approximation, despite
the fact that this was introduced at the beginning of the paper in
order to define time averages. These are the reasons why there is only
one set of equations in \cite{BrillHartle} mixing different orders and
a complete solution to the geon problem is not provided. It is natural
to ask if such a solution based on a spherical shell of nonvanishing
thickness is viable. This question will be answered in the next
Section.

\section{Resolving the geon problem}

In this Section we study the geon problem assuming the high frequency
approximation, as required, and we take into account the orders of
magnitude accordingly. In what follows, we solve the geon problem in
the case of a spherically symmetric, static and asymptotically flat
background $\gamma_{\mu\nu}$.  We solve simultaneously (using
numerical methods) the Einstein equations to the two lowest orders for
a pair $\left( \gamma_{\mu\nu}, h_{\mu\nu} \right)$, taking into
consideration the boundary conditions (\ref{17}). We do not assume
that the metric perturbations vanish outside a certain radius, but
rather solve the system of equations throughout space. We consider odd
modes only and specialize to $m=0$. This is analogous to
Wheeler's electromagnetic geon.  At the end of this Section, the
results will be generalized to $\gamma_{\mu\nu}$ being a
time--dependent, slowly varying, spherically symmetric background
metric.

\subsection{Odd modes}

\setcounter{equation}{0}

To avoid problems which can arise in non--linear equations using a
complex exponential to describe the time dependence of the metric
perturbations, it is advantageous to rewrite the individual modes of
the RW spherical harmonics in their real
form\footnote{Ref.~\cite{geonletter} used $T_0 = T_1 = \cos(\omega
t)$, which is not the most general case. The functions $T_0 $ and $T_1$ are
constrained by the (0,3) component of the Einstein equations.} for $m=0$:
\be  \label{eq4.1}
h_{\mu\nu}^{( \mbox{odd})}\left( t,r,\theta\right)=
\left(
\begin{array}{ccccc}
0 & 0 & 0 & h_0(r) \,\Theta^l(\theta) T_0(t) & \\
                &        &   &    & \\
0 & 0 & 0 & h_1(r) \,\Theta^l(\theta) T_1(t) & \\
                &        &   &    & \\
\mbox{Sym} & \mbox{Sym} & 0 & 0 & \\
                &        &   &    & \\
\mbox{Sym} & \mbox{Sym} & 0 &  0 & \\
\end{array} \right) \; , 
\ee
where
\be
\Theta^l(\theta) = C^{l0} \sin \theta \frac{d}{d\theta} P^l(\cos\theta) \; ,
\ee
\be
T_0(t) = \cos(\omega t + \delta) \; ,
\ee
\be
T_1(t) = \sin(\omega t + \delta) , \ \ \delta \mbox{ = constant} \; .
\ee
The phase constant $\delta$ can be set to zero without loss of generality,
because the phase dependence no longer exists upon time averaging.

The Ricci tensor is computed using Eq.~(\ref{40bis}) which, to the
dominant order O$(1/\epsilon)$, is simplified to (see
Appendix~C)
\be  \label{eq4.2}
R^{(1)}_{\alpha\beta}\left[ \epsilon^{-1}\right]=\frac{1}{2} \,
\gamma^{\rho\tau}\left( h_{\rho\tau,\alpha\beta} + 
h_{\alpha\beta,\rho\tau}-h_{\tau\alpha,\beta\rho}
-h_{\tau\beta,\alpha\rho}\right) \; .
\ee
Substituting Eq.~(\ref{eq4.1}) into Eq.~(\ref{eq4.2}) yields the three
non-trivial equations
\be \label{eq4.3}
h''_0(r) - \omega h_1(r) - l^2 r^{-2} e^\lambda h_0(r) = 0 \; ,
\ee
\be \label{eq4.4}
\omega h'_0(r) + \left( l^2 r^{-2} e^\nu - \omega^2 \right) h_1(r) = 0 \; ,
\ee
\be \label{eq4.5}
h'_1(r) + \omega e^{\lambda - \nu} h_0(r) = 0 \; ,
\ee
from $(\alpha,\beta) = (0,3)$, (1,3) and (2,3)
respectively. Eliminating 
$h_0(r)$ from Eq.~(\ref{eq4.4}) and (\ref{eq4.5}) yield the second
order radial wave equation
\be \label{eq4.6}
h''_1(r) + e^\lambda\left( \omega^2 e^{-\nu} - \frac{l^2}{r^2} \right)
h_1(r) = 0
\ee
for the radial function $h_1(r)$. It should be noted that in the high
frequency limit, the angular momentum quantum number $l \gg 1$, hence
Eq.~(\ref{31}) takes the form
\be   \label{eq4.7}
\frac{d^2 \Theta^l}{d \theta^2} + l^2 \, \Theta^l=0  
\ee
and is used in obtaining Eqs.~(\ref{eq4.3})--(\ref{eq4.6}) (note that
$\cot\theta d\Theta^l /d \theta $ in Eq.~(\ref{31}) is of higher
order than the retained terms in Eq.~(\ref{eq4.7}) in the high
frequency limit).

The next step is to determine the order $O(1)$ equations. Instead of
evaluating Eq.~(\ref{239c}), we can equivalently evaluate 
\be \label{eq4.8}
R^{(0)}_{\mu\nu}(\gamma) -\frac{1}{2}\gamma_{\mu\nu} R^{(0)}(\gamma) =
-\left\langle R^{(2)}_{\mu\nu}\left[\epsilon^0\right]
-\frac{1}{2}\gamma_{\mu\nu}
R^{(2)}\left[\epsilon^0\right]\right\rangle \; .
\ee
By defining the Brill--Hartle space-time averaged stress--energy tensor
as
\be
T^{\mbox{\tiny BH}}_{\mu\nu} = \left\langle T_{\mu\nu}\right\rangle \equiv
-\frac{1}{8\pi} \left\langle R^{(2)}_{\mu\nu}\left[\epsilon^0\right]
-\frac{1}{2}\gamma_{\mu\nu}
R^{(2)}\left[\epsilon^0\right]\right\rangle = -\frac{1}{8\pi}
\left\langle G^{(2)}_{\mu\nu}\left[\epsilon^0\right] \right\rangle \,
, 
\ee
Eq.~(\ref{eq4.8}) takes the familiar form
\be \label{eq4.10}
G^{(0)}_{\mu\nu}(\gamma) = 8\pi T^{\mbox{\tiny BH}}_{\mu\nu} \; ,
\ee
of an effective stress-energy tensor generating the background metric
$\gamma_{\mu\nu}$. 

The procedure for finding the average over the angular dependence of
$T_{\mu\nu}$ has
been given by Wheeler (see ref.~\cite{Wheeler55}, p.~520). The results
are directly applicable to the gravitational geon since the
discussion covers general $T_{\mu\nu}$.  Hence, we have for the
time-angle average (denoted $\langle\rangle_{\mbox{\tiny TA}}$) of
$T_\mu^\nu$ summed over all N active modes 
\be \label{eq4.15b}   
\left\langle 8 \pi
T_0^0\right\rangle_{\mbox{\tiny TA}} = -\frac{N}{2} \int \left\langle
G^{(2)\;0}_{({\mbox{\tiny I}})\;0}\right\rangle_{\mbox{\tiny T}}
\sin\theta \, d\theta         \; ,
\ee 
\be 
\left\langle 8 \pi T_1^1\right\rangle_{\mbox{\tiny TA}} = -\frac{N}{2} 
\int \left\langle G^{(2)\;1}_{({\mbox{\tiny
I}})\;1}\right\rangle_{\mbox{\tiny T}} 
\sin\theta \, d\theta             \; ,
\ee
\be \label{eq4.15a}
\left\langle 8 \pi T_2^2 \right\rangle_{\mbox{\tiny TA}}
=  \left\langle 8 \pi 
T_3^3 \right\rangle_{\mbox{\tiny TA}} \nonumber \\ 
 =  -\frac{N}{2} \int
\left\langle G^{(2)\;2}_{({\mbox{\tiny I}})\;2} +
G^{(2)\;3}_{({\mbox{\tiny I}})\;3}\right\rangle_{\mbox{\tiny T}}
\sin\theta \, d\theta                 \; ,
\ee
\be
\left\langle 8 \pi T_\mu^\nu\right\rangle_{\mbox{\tiny TA}} = 0 \
\mbox{for} \ \mu \neq \nu    \; .
\ee
Here $\langle\rangle_{\mbox{\tiny T}}$ denotes a time average,
and $G^{(2)\;\mu}_{({\mbox{\tiny I}})\;\nu}$ is mode number I of the
disturbance under discussion. Using the mixed form of Eq.~(\ref{R2})
in conjunction with Eqs.~(\ref{eq4.1}) and (\ref{eq4.5}), and
performing the time and angle averaging (see Appendix~D), we obtain
\begin{eqnarray}
 & & \left\langle 8 \pi T_0^0\right\rangle_{\mbox{ \tiny TA}} =
-\frac{N \left(C^{l0} \right)^2 
l}{8 r^2 e^\lambda} \left[-\frac{3}{2}\omega^2 h_1^2 e^{-\nu} -
e^{-\lambda}\left(2 h_1 h_1'' - h_1'^2\right) \right. \nonumber \\
& & \mbox{}- \left. \omega^{-2} e^{-\nu -2\lambda} \left(\frac{1}{2}
h_1''^2 + h_1' h_1'''\right) + \frac{1}{2} l^2 r^{-2}\left(\omega^{-2}
e^{\nu - \lambda}h_1'^2 + h_1^2\right)\right] \; , \label{eq4.14}
\end{eqnarray}
\begin{eqnarray}
 & & \left\langle 8 \pi T_1^1\right\rangle_{\mbox{\tiny TA}} =
-\frac{N \left(C^{l0}\right)^2 
l}{8 r^2 e^\lambda} \left[\omega^{-2} e^{\nu-2\lambda} \left(h_1'
h_1'''-\frac{1}{2} h_1''^2 \right) + e^{-\lambda} h_1'^2 \; .
\right. \nonumber \\ 
 & & \mbox{} \left. + \frac{1}{2}
e^{-\nu}\omega^2 h_1^2 - \frac{1}{2} l^2 r^{-2}\left(\omega^{-2}
e^{\nu - \lambda}h_1'^2 + h_1^2\right)\right] \; . \label{eq4.15}
\end{eqnarray}
We see from Eq.~(\ref{eq4.15a}) that the (2,2) field
equation must be identical to the (3,3) equation. Neither are
necessary to solve the geon problem since the complete system of
equations must be self consistent.  The latter two equations may
be derived from the $(1,1)$, $(0,0)$ and wave equations. It is
of interest to compare Eqs.~(\ref{eq4.14})--(\ref{eq4.15}) with their
electromagnetic counterparts. The following equations are given in
Ref.~\cite{Wheeler55} and are not restricted to the high frequency
approximation:
\be
\left\langle 8 \pi T_1^1\right\rangle_{\mbox{\tiny TA}} =
\frac{N\, l(l+1)}{2(2l+1)}\left[ e^{-\lambda}
\left(\frac{dR}{rdr}\right)^2  
+  e^{-\nu} \left(\frac{\Omega R}{r}\right)^2 -l(l+1)\left(\frac{
R}{r^2}\right)^2 \right]   \; ,
\ee
\be
\left\langle 8 \pi T_0^0\right\rangle_{\mbox{\tiny TA}} =
\frac{N\, l(l+1)}{2(2l+1)}\left[ -e^{-\lambda}
\left(\frac{dR}{rdr}\right)^2  
-  e^{-\nu} \left(\frac{\Omega R}{r}\right)^2 -l(l+1)\left(\frac{
R}{r^2}\right)^2 \right]       \; .
\ee
Here $R=R(r)$ is the electromagnetic counterpart of $h_1(r)$. It is
evident that the similar terms (same differential order) in $T^0_0$ for
the gravitational case have a sign difference. For the $T^0_0$
component, one of the three similar terms has a sign
difference. These sign differences play an important role in
the subsequent solving of the gravitational geon problem. It will be
shown that the sign differences lead to a negative mass for the
gravitational geon as the only non--trivial solution to the field
equations. It is understandable that these sign differences should
arise: in the electromagnetic case, the source is designed at will 
(subject to the Maxwell equations) and
appears on the right hand side of the Einstein equations. By contrast,
the gravitational source is artificially constructed from terms on the
left hand side of the Einstein equations which are shifted to the
right hand side.

The left hand side of Eq.~(\ref{eq4.10}) is 
\be \label{eq4.16}
G^{(0)\;0}_{\;\;\;\;0} = e^{-\lambda}\left(r^{-2} - r^{-1}\lambda'\right)
- r^{-2}      \; ,
\ee 
\be \label{eq4.17} 
G^{(0)\;1}_{\;\;\;\;1} =
e^{-\lambda}\left(r^{-2} + r^{-1}\nu'\right) - r^{-2} \; .
\ee
The equations for the proposed gravitational geon have the same scale
invariance as those for the electromagnetic counterpart. Therefore we
can introduce the same dimensionless measure of radial coordinate,
$\rho = \omega r$. By analogy with the electromagnetic case, we can
define a new measure of potential
\be
f(\rho) = \sqrt{k^l}\,\omega h_1(r), \mbox{ where } k^l \equiv
\frac{1}{8} \, l N \left(C^{l0} \right)^2 \; ,
\ee
and two metric functions $L(\rho)$ and $Q(\rho)$ such that
\be \label{eq4.18}
e^{-\lambda} \equiv 1 - 2\rho^{-1}L(\rho) \; ,
\ee 
\be \label{eq4.19}
e^{\lambda+\nu} \equiv Q^2(\rho) \; ,
\ee 
\be \label{eq4.20}
e^{\nu} = \left[1 - 2\rho^{-1}L(\rho)\right]Q^2(\rho) \; .
\ee 
Substituting the above into
Eqs.~(\ref{eq4.6}), (\ref{eq4.14})--(\ref{eq4.17}) yelds the
wave equation
\be \label{eq4.21}
\frac{d^2f}{d\rho^{*2}} + \left[ 1 -\left(\frac{l Q}{\rho}\right)^2
\left(1-\frac{2L}{\rho}\right)\right] f = 0     \; , 
\ee
where
\be \label{eq4.22}
d\rho^* = Q^{-1}\left(1-\frac{2L}{\rho}\right)^{-1}d\rho  \; , 
\ee
and the two background equations
\begin{eqnarray}
 & & \frac{dL}{d\rho^*} = -\, \frac{\left(1-2L/\rho\right)}{2Q} \left[
\frac{3}{2} f^2 + 2 f \frac{d^2f}{d\rho^{*2}} +
\left(\frac{df}{d\rho^*}\right)^2 \right. \nonumber \\
& & \mbox{} + \frac{1}{2}\left.\left(\frac{d^2f}{d\rho^{*2}}\right)^2 +
\frac{df}{d\rho^*}\frac{d^3f}{d\rho^{*3}}  - \frac{1}{2}\rho^{-2}
l^2Q^2\left(1-2L/\rho\right) \left(\left(\frac{df}{d\rho^*}\right)^2
+ f^2\right) \right] \; , \label{eq4.23}
\end{eqnarray}
\begin{eqnarray}
& & \frac{dQ}{d\rho^*} =-\, 
\frac{\left(1-2L/\rho\right)}{\rho-2L} \left[ f^2 + f \frac{d^2f}{d\rho^{*2}}
+ \left(\frac{df}{d\rho^*}\right)^2 \right. \nonumber \\
& & \mbox{} + \left.\frac{df}{d\rho^*}\frac{d^3f}{d\rho^{*3}}
- \frac{1}{2}\rho^{-2}
l^2Q^2\left(1-2L/\rho\right) \left(\left(\frac{df}{d\rho^*}\right)^2
+ f^2\right) \right] \; . \label{eq4.24}
\end{eqnarray}
As in the electromagnetic case, the above equations permit a further
change of scale without change of form:
\be \label{eq4.25}
\rho = b\rho_1,\ Q=bQ_1,\ \rho^* = \rho^*_1 ,\ L=bL_1,\
f=b^{\frac{1}{2}} f_1 \; .
\ee
This allows $Q_1(0)\equiv 1$ and then the scaling parameter can be
found by demanding $Q(\infty)~=~1$. Note that in Eqs.~(\ref{eq4.23})
and (\ref{eq4.24}) (and in Wheeler's electromagnetic study), an
average over a length larger than the characteristic wavelength of the
radial function $f(\rho)$ has {\em not} been performed. This would
enormously complicate the system of equations (\ref{eq4.21})--(\ref{eq4.24})
and was not believed to be necessary in Ref.~\cite{Wheeler55}. Our approach
parallels that of Ref.~\cite{Wheeler55}.

Since the wave equation for the proposed gravitational geon is
identical to the one for the electromagnetic counterpart, the same
arguments hold for applying another scaling law valid asymptotically
for large $l$ (the high frequency of the gravitational waves
guarantees large $l$). By making the transformation
\be \label{eq4.26}
x = \left(\rho^* - l\right) l^{-1/3},
\ee
the entire active region of the proposed geon will be described by a
range of $x$ of order unity. Wheeler \cite{Wheeler55} has provided the
expansion of the relevant quantities in inverse powers of
$l^{1/3}$. They are
\begin{eqnarray}
d\rho^* & \equiv & l^{1/3} dx \; , \nonumber \\
\rho_1 & = & l + l^{1/3} r_0(x) + \cdots \; , \nonumber \\
L_1 & = & l \lambda_0(x) + l^{2/3} \lambda_1(x) + l^{1/3} \lambda_2(x)
+ \cdots \; , \label{eq4.27} \\
Q_1 & = & 1/k(x) + l^{-1/3} q_1(x) + l^{-2/3} q_2(x)
+ \cdots \; , \nonumber \\
f_1 & = & l^{1/3} \phi(x) +  \phi_1(x) + l^{-1/3} \phi_2(x)
+ \cdots \; . \nonumber 
\end{eqnarray}
After substituting Eq.~(\ref{eq4.27}) into (\ref{eq4.21}), expanding in
inverse powers of $l^{1/3}$ and setting the lowest three orders to
zero, we find the two algebraic equations
\be \label{eq4.28}
\lambda_0 = \frac{1}{2}\left( 1 - k^2\right),
\ee
\be \label{eq4.29}
\lambda_1 = q_1 k^3,
\ee
and the differential equation
\be \label{eq4.30}
\frac{d^2\phi}{dx^2} + j(x) k(x) \phi(x) = 0 \; .
\ee
Here, we have defined 
\be \label{eq4.31}
j(x) \equiv \frac{1}{k^3}\left(-r_0-2q_2 k^3 + 2
\lambda_2+3q_1^2k^4+3k^2 r_0\right)
\ee
after using Eqs.~(\ref{eq4.28}) and (\ref{eq4.29}). Repeating this
procedure for Eq.~(\ref{eq4.23}), we obtain
\be \label{eq4.32}
\frac{dk}{dx} = \frac{1}{2} k^2 \phi^2 \; ,
\ee
\be \label{eq4.33}
\frac{dq_1}{dx}= - \phi \phi_1    \; ,
\ee
\begin{eqnarray}
& & \frac{d\lambda_2}{dx} + \frac{1}{4} k^3 r_0 \phi + \frac{9}{4} q_1^2
k^5 \phi^2 - 3q_1 k^4 \phi \phi_1 + \frac{1}{2} k^3 \phi_1^2 +
\frac{1}{4} k^3 \left(\frac{d\phi}{dx}\right)^2\nonumber 
\\
& & \mbox{} + \phi\phi_2 k^3 + k^3 \phi \frac{d^2\phi}{dx^2} - q_2 k^4
\phi^2 - \frac{1}{2} \lambda_2 k \phi^2 + \frac{1}{4} k r_0 \phi^2 = 0 \; .
\label{eq4.34} 
\end{eqnarray}
The expansion of Eq.~(\ref{eq4.24}) yields Eqs.~(\ref{eq4.32}),
(\ref{eq4.33}) and 
\begin{eqnarray}
& & \frac{dq_2}{dx} + r_0 \phi^2 - q_2 k \phi^2 +\frac{1}{2} \phi_1^2
 + \lambda_2 k^{-2} \phi^2 +\frac{1}{2} 
\left(\frac{d\phi}{dx}\right)^2 \nonumber \\
& & \mbox{} + \phi \phi_2
+ \phi \frac{d^2\phi}{dx^2} +\frac{3}{2} q_1^2 k^2 \phi^2 -\frac{1}{2}
r_0 \phi^2 k^{-2} = 0   \; .
\label{eq4.35} 
\end{eqnarray}
By differentiating Eq.~(\ref{eq4.31}), utilizing Eqs.~(\ref{eq4.28}),
(\ref{eq4.29}), (\ref{eq4.32}), (\ref{eq4.33})--(\ref{eq4.35}) and
substituting 
\be \label{eq4.36}
\frac{dr_0}{dx} = k
\ee
(derived from Eqs.~(\ref{eq4.22}) and (\ref{eq4.27})), we obtain the
differential equation
\be \label{eq4.37}
\frac{dj(x)}{dx} = 3 - \frac{1}{k^2}\left[ 1 - \frac{1}{2} k^2
\left(\frac{d\phi}{dx}\right)^2 \right] \; .
\ee
Solving Eqs.~(\ref{eq4.30}), (\ref{eq4.32}) and (\ref{eq4.37})
simultaneously for the three functions $\phi(x)$, $j(x)$ and $k(x)$ is
sufficient for determining the remaining leading terms in
Eq.~(\ref{eq4.27}). It should be noted that since Eqs.~(\ref{eq4.30}),
(\ref{eq4.32}) and (\ref{eq4.37}) do not explicitly depend on $x$, the
system of equations is autonomous. Hence, if $\phi(x),\, j(x)$ and
$k(x)$ are solutions , then so are $\phi(x+a),\, j(x+a)$ and $k(x+a)$
where $a$ is a constant. Thus the gravitational geon problem is
reduced to finding a solution to the system
\be \label{ggwave}
\frac{d^2\phi}{dx^2} + j(x) k(x) \phi(x) = 0 \; .
\ee
\be \label{ggk}
\frac{dk}{dx} = \frac{1}{2} k^2 \phi^2   \; ,
\ee
\be \label{ggj}
\frac{dj(x)}{dx} = 3 - \frac{1}{k^2}\left[ 1 - \frac{1}{2} k^2
\left(\frac{d\phi}{dx}\right)^2 \right]      \; ,
\ee
with the following properties (which are identical to the conditions
for the electromagnetic geon problem):
\begin{enumerate}
\item{\em For large negative $x$:} The field factor
$\phi(x) \rightarrow 0$ and $k(x)\rightarrow 1$. Under these
conditions $dj(x)/dx = 2$ or $j(x) = 2 x$. Choosing the integration
constant for $j(x) = 2 x$ to be zero fixes $a$ and consequently
defines the origin of $x$. This removes any ambiguity in the start of
the integration process. Thus for large negative
$x$, $\phi(x)$ satisfies the equation
\be
\frac{d^2\phi}{dx^2} = 2 x \phi(x) \ .
\ee
The approximate solution as given by Wheeler \cite{Wheeler55} is 
\be
\phi(x) \equiv \frac{A}{3} \left(-2x\right)^{-1/4} 
\exp[-(-2x)^{3/2}]    \; .
\ee
\item{\em For large positive $x$:} It is required that $\phi(x)
\rightarrow 0$, $0 < k(x) < 1$ and $j(x)$ approach large negative
values. 
\end{enumerate}

The only free parameter is the amplitude $A$ of the wave and this must
be chosen so that the solution fits the boundary conditions. The
non-linearity of the problem makes it necessary to integrate the
system of equations numerically. The integration is started at
$x=-4$. The initial conditions are as follows:
\be \label{eq4.40}
\phi(-4) = \phi_0 = \mbox{ arbitrary } \; ,
\ee
\be
\left. \frac{d\phi}{dx} \right|_{x=-4} = \left(\frac{1}{16} +
\sqrt{8}\right) \phi_0          \; ,
\ee
\be
k(-4) = 1   \; ,
\ee
\be \label{eq4.43}
j(-4) = -8      \; .
\ee
Figure~1 shows the result of the numerical integration using
a variable step fourth-fifth order Runge--Kutta method. 

In order to understand the significance of the curves plotted in
Fig.~1, it would be beneficial to briefly review some of the
properties of the proposed electromagnetic geon solution. The
differential equations for the electromagnetic case are
\be \label{eq4.44}
\frac{d^2\phi}{dx^2} + j(x) k(x) \phi(x) = 0 \; ,
\ee
\be \label{eq4.45}
\frac{dk}{dx} = - \phi^2     \; ,
\ee
\be \label{eq4.46}
\frac{dj(x)}{dx} = 3 - \frac{1}{k^2}\left[ 1 +
\left(\frac{d\phi}{dx}\right)^2 \right] \; .
\ee
The initial conditions for these equations are
given by Eqs.~(\ref{eq4.40})--(\ref{eq4.43}) at $x=-4$.

The ``active region'' of the geon is defined to be the range of $\rho$
where the square bracketed combination of terms in Eq.~(\ref{eq4.21})
is positive. In this region, the function $f(\rho)$ has oscillatory
behaviour. Where the bracketed terms are negative, the behaviour of
$f(\rho)$ is exponential growth or decay. The active region can be
identified in the $x$ coordinate system as the region where the
oscillating factor $j(x) k(x)$ is positive. The mass of the geon
inside radius $\rho$ is related to the function $k(x)$ in the
following way:
\be M(\rho(x)) = \frac{1}{b} \lambda_0(x) =
\frac{1}{2b} \left(1-k^2\right)     \; ,
\ee 
with $b=1/Q_1(\infty) = k(\infty)$. This implies that 
\be 
0 \le k(x) \le 1 \mbox{ or }  \mbox{ as } x \rightarrow \infty 
\ee 
in order to have a positive total mass.

An eigenvalue solution for Eqs.~(\ref{eq4.44})--(\ref{eq4.46}) is
sought for which $\phi(x) \rightarrow 0$ as $ x \rightarrow
\infty$. If the amplitude factor $A$ (which translates into an initial
choice of the field factor $\phi_0$) is slightly higher than the
desired eigenvalue, $\phi(x)$ reaches a minimum and then rises
exponentially and becomes singular at some finite $x$. For an
amplitude slightly less than the eigenvalue, $\phi(x)$ goes to
$-\infty$ at a finite $x$.

Figures~2 and~3 show that the first eigenvalue (characterized by
$\phi(x)$ having one maxima and no local minima) appears to lie
between those amplitude factors $A$, that correspond to initial
values of $\phi_0$ at some point in the range $ 9.790419489\times
10^{-5} < \phi_0 < 9.790419490\times 10^{-5}$. The mass factor $k(x)$
give a positive mass throughout the integrable region (before the
singularity) and appears to have a $k(\infty)$ value of approximately
1/3. The function $j(x)$ is positive only for a limited range in the
neighbourhood of $x=1$, thus identifying the active
region. Qualitatively, these results are similar to Wheeler's
computations. The only main difference between Wheeler's calculation
and the present one is that his first eigenvalue lies at some point in
the range $1.03000\times 10^{-4}< \phi_0 <1.03125\times 10^{-4}$ and
the active region starts at $x=4.05$ and ends at $x=6.02$.

For the proposed gravitational geon, the behaviour of the functions is
quite different from that of the electromagnetic case. This is evident
in the differential equations themselves. Comparing
Eqs.~(\ref{eq4.45}) and (\ref{ggk}), there is a sign difference on the
right hand side of the equations. This is the manifestation of the
sign difference identified earlier in the comparison of the
electromagnetic and gravitational stress--energy tensors. Proceeding
with the analysis of the equations, Fig.~1 shows the results of the
integration of $\phi(x)$, $k(x)$ and $j(x)$ for the initial value
$\phi_0 = 4.45\times 10^{-4}$. The factor $j(x) k(x)$ becomes positive
at approximately $x=0$ and singular at approximately $x=1$. Since
$j(x) k(x)$ never becomes negative again, it implies that there is no
end to the active region and as a consequence, $\phi(x)$ remains
oscillatory for large $x$. Of an even more disturbing nature, the
function $k(x)$ appears greater than 1 for all $x$ and approaches
$+\infty$. The implication of this is that the mass of the proposed
gravitational geon is negative. As $\phi_0$ is decreased (increased),
the singular behaviour of $k(x)$ and $j(x)$ moves to larger (smaller)
values of $x$, but $k(x)$ remains greater than 1 and $j(x) > 0$ once
it becomes positive. Apparently, the only physical solution would be
for $\phi_0 =0$, which yields $k(x) =1$ and hence a zero mass
gravitational geon. Thus we conclude that it is not possible to
construct a gravitational geon and the only physical solution is the
trivial solution.

\subsection{The time--dependent and stationary cases}

The previous results can be generalized to the case of a
time--dependent, spherically symmetric background metric
$\gamma_{\mu\nu}\left( t,r \right)$, under the assumption that its
time variation occurs on a scale much larger than the period of the
gravitational waves. In this case the high frequency approximation and
Eqs.~(\ref{8})--(\ref{10bis}) remain valid. Equation~(\ref{2}) still
holds, but Eq.~(\ref{3}) is replaced by
\be                   \label{81}
\lambda=\lambda(t,r) \; , \;\;\;\;\;\;\;\;\;\; \nu=\nu(t,r) \; .
\ee
As a consequence of the fact that the estimate of the orders of magnitude
in the Einstein equations does not change, we find in this case the same
equations that were presented above for the odd modes, and the
same conclusions apply. If instead, the background metric
$\gamma_{\mu\nu}\left( t, r \right) $ is allowed to vary on a time
scale comparable to the period of the gravitational waves, the high
frequency approximation does not hold and a gravitational geon cannot
be constructed, as explained in Sec.~2. This remains valid for any
time--dependent background metric $\gamma_{\mu\nu} \left( t,
\vec{x}\right)$ when symmetries are absent, due to the fact that our
considerations based on Eq.~(\ref{239c}) do not rely on the assumption
of spherical symmetry. Apart from this argument, the
realization of a geon with a rapidly varying background metric
$\gamma_{\mu\nu}$ is problematic for another reason: If a spherically
symmetric background is allowed to vary
harmonically with frequency $\Omega$ comparable to the frequency of
the gravitational waves, one expects a parametric resonance
\cite{Arnold} for the modes with $\omega =n \Omega/2$, with
$n=1,\,2,\,\cdots$~. The strength of the resonance is a maximum for
$n=1$ and decreases rapidly as $n$ increases. In the limit of a static
background, the resonance phenomenon disappears. Accordingly, on
the basis
of studies of perturbations of black holes and relativistic stars
\cite{Chandrasekhar}, it is expected that in the case of a
stationary axisymmetric background metric describing a rapidly
rotating geon,
the resonance phenomenon between the perturbations and the background
metric occurs.
In the general case of a time--dependent and rapidly varying background
metric $\gamma_{\mu\nu}\left( t, \vec{x}\right)$ without symmetries, it
is not known how to decompose metric perturbations on a complete set
playing the role of the tensor spherical harmonics in the spherical
case, or even how to define frequencies in the strong curvature
region. However, if such concepts can be given a meaning, it seems
reasonable to expect some kind of resonance phenomena between the
background
metric and its gravitational wave perturbations. All these resonance
phenomena certainly do not contribute to the realization of a stable
configuration, but rather are associated with instabilities that
tend to disrupt the system.

\section{Other approaches to the geon problem}

In the previous sections, we have analyzed the BH construct and the models of
gravitational geons conceived by Wheeler. However, one can study different
models of a gravitational geon and different, independent, approaches to the
geon problem, which are given in the present Section.

A first argument, which provides additional intuitive physical 
insight, is the following: We recall our
analogy of Sec.~2 between gravitational waves composing a geon and
stars composing a galaxy. The high frequency approximation required in
the geon case has a parallel in the case of a galaxy; it corresponds
to the requirement that the individual stars have a very high
velocity. It is clear that such stars would escape from the galaxy and
would not be trapped by its potential well. A galaxy cannot be built
exclusively from such stars in rapid motion. In other words, the
system would not satisfy the virial theorem and would not be
bounded. The difference with the gravitational geon case is that while
one is not obliged to require that stars have a very high velocity
when constructing a galactic model, the high frequency approximation
is necessary for a geon and this, in turn, prevents its realization.

An independent argument to understand the impossibility of a
gravitational geon is the following: it is well known that, in the
limit of high frequencies, gravitational waves obey the laws of geometric
optics \cite{Isaacson,MTW}. Spatially closed lightlike
geodesics exist only inside black holes, which necessitate the
existence of singularities. Thus, they are necessarily inconsistent
with the definition of a geon. The null circular geodesic at $r=3M$ in
the Schwarzschild geometry is unstable. It is therefore hard to
reconcile high frequency gravitational waves with stable trapped
graviton trajectories in the absence of matter.

The most intuitive model of a gravitational geon is that of a ball of
high frequency gravitational radiation, which behaves like 
a perfect fluid with
a radiation equation of state. It appears that Wheeler was aware of this
possibility, but discarded it as non--viable and therefore proceeded to 
study the more
complicated models of Ref.~\cite{Wheeler55}, in which the waves do not
propagate radially and the radiation is not
isotropic. ``...~one naturally recalled that some stars derive their energy
almost exclusively from particles; others, from a mixture of particles and
radiation. The extreme limit of a system deriving its mass--energy from
radiation alone therefore suggested itself. However, with no matter to provide
opacity and to dam up the radiation against escape, stability could only be
maintained by excluding all photon orbits in which the motion is purely radial
or largely radial~...'' \cite{HTWW}. However plausible this argument may 
appear, it 
relies on the theory of the stability of Newtonian stars, and cannot 
be used for a 
relativistic fluid ball, since in general relativity the fluid pressure 
contributes to the energy density which may bind the fluid and this
contribution cannot be neglected in a relativistic star dominated by
radiation pressure. Therefore, it is important to re--examine 
the possibility of
a fluid ball made only of gravitational waves. There have been 
some papers (see 
\cite{Sorkinetal,Sokolov} and
references therein) analyzing the properties of self--gravitating
electromagnetic radiation confined to a spherical box. The radiation
is taken to be a perfect fluid with equation of state $p=\rho/3$. This
model is as applicable to high frequency gravitational waves as it is to 
electromagnetic waves, hence the
results of these papers will hold true for gravitational radiation. In
order to build a geon (both electromagnetic and gravitational), the
constraint of the spherical box (with reflecting walls) would have to
be removed. This requirement leads to the impossibility of
constructing any type of geon using a relativistic perfect fluid
model. Weinberg \cite{Weinberg} has shown that a highly relativistic
fluid with $p=\rho/3$ can never achieve hydrostatic equilibrium in a
finite ball through gravity alone. The relativistic equation for
hydrostatic equilibrium is\setcounter{equation}{0}
\be
- \frac{\partial p}{\partial x^\lambda} = (p+\rho) \frac{\partial}
{\partial x^\lambda} \ln \left[ ( -g_{00} ) ^{1/2}\right] \; .
\ee 
For $p=\rho/3$, $\rho \propto (-g_{00})^{-(p + \rho)/2 p}$. Since
$\rho$ must vanish outside the fluid, $g_{00}$ would have to become
singular at its surface. If the surface of the ball is allowed to
extend to spatial infinity, then from Sokolov's work \cite{Sokolov},
it is not difficult to establish that the mass of the radiation ball
diverges (see also Ref.~\cite{MTW}, p.~615, ex. 23.10).

Another work which casts doubt on the existence of electromagnetic
geons is that of Gibbons and Stewart \cite{Gibbons}. They conclude
that the Einstein equations do not permit asymptotically flat
solutions which are both periodic and empty near infinity.  This
precludes the existence of gravitational geons for at least the
periodic case. They also suggest that the result may be extended to
include the case when there is matter near infinity,
e.g. electromagnetic or scalar radiation.

\section{Discussion and conclusion}

The results of the previous sections were derived by making use of some
particular gauge conditions that RW imposed in order to
set the metric perturbations in the form of Eqs.~(\ref{4ter}) and
(\ref{4bis}). However, it is clear from their very nature that our
results are covariant and gauge--independent, since the solution
$\left( \gamma_{\mu\nu},h_{\mu\nu} \right)=\left( \eta_{\mu\nu},0
\right)$ that we found has an invariant meaning (for example, the
vanishing of the curvature tensor is a covariant concept).

Since a spherically symmetric gravitational geon cannot exist due to
the fact that the high frequency approximation does not allow a
solution with the required characteristics, one might ask if it is
possible to realize a gravitational geon in a configuration with less
symmetry.  We do not expect that such a geon can be constructed when
the most primitive case is excluded. The main reason for this belief
is that the key factor which leads to the non--existence of the
spherical geon is not the spatial symmetry but rather the high
frequency.

From a mathematical point of view, the main difference between our
approach to the geon problem, as compared to that of BH, consists in
our explicit use of the high frequency approximation in conjunction
with solving explicitly for the wave and background metric functions
in a self--consistent manner. We have already seen in Sec.~2 that this
is necessary for the geon problem to be meaningful. In Sec.~4 it was
shown that the same approximation prevents the realization of a
spherically symmetric geon.

In his papers on geons, Wheeler
\cite{Wheeler55,PowerWheeler,Wheeler61,Wheeler62} describes
electromagnetic and neutrino geons as systems which are stable on a
long time scale, but not absolutely stable, in the sense that they
``leak'' radiation to the exterior. The rate of the leaking is
negligible, so that a geon is stable for a long period of
time. However a secular instability is introduced, which seems
unavoidable \cite{Wheeler61}. The BH model of a spherical shell with
$h_{\mu\nu}$ exactly equal to zero outside a certain radius excludes
such a possibility, and it could be conjectured that this might be the
reason why their model is not viable, leaving a possibility open for
the realization of physically more realistic ``leaking''
geons\footnote{There is inconsistency in \cite{BrillHartle} at this
point: in that paper it is required that $h_{\mu\nu}$ (and therefore
$Q$) vanishes outside the spherical shell. However, the
Schr\"{o}dinger--like equation that is derived there for $Q$ (our
Eq.~(\ref{39})) implies a ``leaking'' geon, as is stated in
\cite{BrillHartle}. In fact, the function $Q$ has a nonvanishing tail
for large values of the radius, due to the fact that the effective
potential barrier is finite. This effect is analogous to the
well--known tunnel effect in quantum mechanics.  }. However, this
possibility is excluded by our calculations.  In fact our boundary
conditions (\ref{17}) allow for this possibility, which in turn is
excluded by our results as well.

It should be noted that while there is a marked contrast between the
characteristics of proposed gravitational and electromagnetic geons as
evidenced in Figs. 1, 2 and 3, it is still unclear that even an
electromagnetic geon is a viable entity. This is because the
electromagnetic geon plots do not display true soliton--like
confinement. Rather they show a region which, by very high refinement
of the eigenvalue, is very near vacuum but which is followed by an
infinite amount of energy beyond this region. It was
contended~\cite{Wheeler55} that by continued refinement of the
eigenvalue, this infinite energy region would itself be pushed out to
infinity. Firstly, it is not clear that this really is the case and it
is possible that infinite refinement of the eigenvalue could lead to a
convergence of the infinite energy regime at some {\em finite}
distance from the vacuum region. Such would clearly be unacceptable as
a model of the desired confined energy concentration. Secondly, even
if it is the case as was conjectured, that continued refinement of the
eigenvalue would push the infinite energy regime out to infinity, this
would not appear to be the idealized soliton--like structure that is 
being sought.

Traditionally, the geon was conceived as a structure of
small--amplitude high--frequency gravitational waves compactified to
the point where one could describe the resulting metric as the
averaged ``background'' metric induced by the totality of the waves
plus a small perturbation due to the local wave presence.  This is
what was analyzed in the present work. It is natural to consider also
waves of ``large'' amplitude in which case linearization is no longer
possible nor is it meaningful to envisage a splitting of the metric as
before. In fact, to assign a measure to amplitude presupposes a
standard for comparison and in the present work, the background metric
served this role. To speak now of large amplitude is to consider waves
for which there is no longer a discernible ``background'' and hence no
standard for comparison of amplitude measure. This leads to the realm
of exact solutions. One might ask whether an exact wave--like solution
of the Einstein equations, singularity--free with localized curvature
and asymptotically flat, could exist. Existing exact wave--like
solutions such as the plane waves of Bondi, Pirani and Robinson or the
cylindrical waves of Einstein and Rosen \cite{KSMH} are not localized
and in the second case, are also not singularity--free. While it would
appear doubtful that solutions with the geon--like properties can
exist, to our knowledge they are not ruled out.

Implicit in the gravitational geon concept is the assumption that the
gravitational field has some particular essential features shared by
other fields. Other fields, even in their pure states, carry
energy. Energy has a mass equivalent and all masses gravitate. Thus,
given a sufficient concentration of field energy, one could imagine a
gravitated concentration into a spherical region with the effective
mass displayed unambiguously by the coefficient of the $1/r$ part of
the asymptotic static vacuum metric.  The gravitational geon concept
is built upon the assumption that the gravitational field itself, even
in its pure state, will gravitate and thus have the potential to
behave as other concentrations of matter or fields.  Through the
years, various authors such as Isaacson \cite{Isaacson} have dwelt
upon the similarities between the gravitational and other fields.  For
example, Isaacson has attempted to establish that there is a basis for
considering a certain construct of the metric as an energy--momentum
tensor of the gravitational field which is as substantial as a true
energy--momentum tensor. However, this requires averaging and under
the appropriate limits, his construct merges with the energy--momentum
pseudotensor, the shortcomings of which epitomize the gravitational
energy problem. If the gravitational field in its pure form really did
have the properties which those authors have ascribed to it, then it
would seem reasonable to expect that a gravitational geon could, at
the very least in principle, be constructed.  However, given the
present results, it is worth considering alternative ideas.

Recently, one of the authors \cite{Cooperstock,Cooperstock2}
introduced a new hypothesis that gravitational energy is localized in
regions of non--vanishing energy--momentum tensor. The motivation
derived from the fact that the traditional means by which physicists
have identified gravitational energy was through the covariant
energy--momentum conservation laws. While those laws were extrapolated
to produce energy--momentum pseudotensors, implying densities and
fluxes even in vacuum, the fact is that the laws themselves are devoid
of content in vacuum, producing the empty identity $0=0$. Given that
there is a plethora of possible pseudotensors and, as their name
implies, they are not really tensors, it was suggested
\cite{Cooperstock} that the root of the ambiguity lies in the
extrapolation of the conservation laws to regions in which they are
without actual content. The hypothesis goes on to propose that the
true expression of the gravitational contribution to energy is
confined to regions of non--vanishing $T_{\mu\nu}$. In a sense this is
the opposite of the Isaacson approach in that rather than being
satisfied with a construct which reduces to the pseudotensor, the new
hypothesis suggests that proper localization is realized when the
pseudotensor is removed.

Clearly, the realization of a gravitational geon would negate the new
hypothesis as it would provide an example of a space totally free of
true energy--momentum tensor $T_{\mu\nu}$ yet exhibit an unambiguous
energy content via its asymptotic metric. While one might propose
exact plane gravitational wave solutions as counter--examples to the
hypothesis, it is to be noted that these are unbounded fields with
questionable relevance to physical situations and more directly, these
wave solutions can be expressed in Kerr--Schild form for which the
pseudotensor vanishes in its entirety \cite{GursesGursey}.  The
gravitational geon is a direct challenge to the hypothesis and if the
geon cannot exist, the hypothesis has passed another test.

\section*{Acknowledgments}

We are grateful to several colleagues for helpful criticisms. 
This research was supported, in part, by a grant from the Natural
Sciences and Engineering Research Council of Canada. 

\section*{Appendix A: Derivation of Eq.~(\ref{31})}

We start from the Legendre
equation\def\theequation{A.\arabic{equation}}\setcounter{equation}{0}
\be   \label{A1}
\frac{d}{dx} \left[ \left( 1-x^2 \right)
\frac{dP^l(x)}{dx}\right]+l(l+1)P^l(x)=0                \ee
and note that
\be            \label{A2}
\Theta^l( \theta)=C^{l0}
\sin \theta \, \frac{dP^l( \cos \theta
)}{d\theta}=C^{l0}
\left( x^2-1\right) \frac{dP^l(x)}{dx}  \; ,          \ee
where $x=\cos \theta$. Using
\begin{eqnarray}
&& \frac{d}{d\theta}=-\sin \theta \, \frac{d}{dx}    \; ,
\label{A3}\\
&& \frac{d^2}{d\theta^2}=\sin^2 \theta \, \frac{d^2}{dx^2}-\cos \theta
\, \frac{d}{dx}        \; ,             \label{A4}
\end{eqnarray}
and the Legendre equation (\ref{A1}), we find the relations
\be              \label{A5}
\frac{d \Theta^l}{d\theta}=-l(l+1) C^{l0}
\sin \theta \, P^l(x)      \; ,           \ee
\be              \label{A6}
\frac{d^2 \Theta^l}{d\theta^2}=-l(l+1) C^{l0}
\left[ xP^l(x)+\left(x^2-1\right)
\frac{dP^l(x)}{dx} \right]    \;  .      \ee
Using Eqs.~(\ref{A5}) and (\ref{A2}) in Eq.~(\ref{A6}), Eq.~(\ref{31})
follows.

\section*{Appendix~B: Junction conditions for the BH background metric}

We consider the Darmois junction conditions \cite{junction} for the BH
background metric on the timelike hypersurface $ S \equiv
\left\{ \left(t,r,\theta,\varphi \right) :\;\;\; r=a \right\}$
separating
the regions of the spacetime manifold $ U\equiv \left\{
\left(t,r,\theta,\varphi \right) :\;\;\; r<a \right\} $,
$\bar{U}\equiv \left\{ \left(t,r,\theta,\varphi
\right) :\;\;\;r>a \right\} $. $ \left\{ x^{\alpha} \right\}=\left\{
\bar{x}^{\alpha} \right\}=\left\{ t,r,\theta,\varphi \right\}$ and
$\left\{ u^i \right\}_{i=0,2,3}=\left\{ t,\theta,\varphi\right\}$ are
coordinate systems in $U$, $\bar{U}$ and $S$, respectively (note that,
in this Appendix, Latin indices assume the values 0,~2,~3 due to the
timelike character of $S$). The unit normal to $S$ is directed along
the coordinate basis vector dual to $dr$ and has components
\def\theequation{B.\arabic{equation}}\setcounter{equation}{0}
\be      \label{II1}
n_{\mu}=\delta^1_{\mu}\, \mbox{e}^{\lambda/2}   \; .
\ee
The metric components $\gamma_{\mu\nu}$ in $U$ and
$\bar{\gamma}_{\mu\nu}$
in $\bar{U}$ are given by Eqs.~(\ref{2}), (\ref{star1}) and
(\ref{star2}). The first fundamental form of $S$ has components $
\gamma_{ij}=\bar{\gamma}_{ij}$. The second fundamental form
$ K_{\mu\nu}
\equiv n_{\mu;\nu}$ of any hypersurface $r=$constant has components
\be                    \label{II2}
K_{ij}=n_{\alpha ;\beta}\,  \frac{\partial x^{\alpha}}{\partial u^i}
\,\frac{\partial x^{\beta}}{\partial u^j}=-\Gamma^1_{ij}\,
\mbox{e}^{\lambda/2}
\ee
in coordinates $\left\{ u^i \right\}$. Using the Christoffel
symbols of a spherically symmetric metric (see e.g.
\cite{LandauLifschitz}), we obtain the only nonvanishing
components
\begin{eqnarray}
&& K_{00}=-\, \frac{\nu'}{2}\,\mbox{e}^{\nu-\lambda/2}
\; , \label{II3} \\
&& K_{22}=r\, \mbox{e}^{-\lambda/2}   	\; ,
\label{II4}     \\
&& K_{33}=r \,\mbox{e}^{-\lambda/2} \sin^2\theta  \; .
\label{II5}
\end{eqnarray}
The Darmois conditions \cite{junction} require the continuity of the
first and second fundamental form across $S$. The first condition is
trivially satisfied, while the second is violated. In fact, we have
\begin{eqnarray}
&& \lim_{r\to a^-} K_{00}=0\neq \lim_{r\to a^+} K_{00}=-\,
\frac{16}{27M} \; ,       \label{II6} \\
&& \lim_{r\to a^-} K_{22}=a \neq \lim_{r\to a^+}
K_{22}=\frac{a}{3} \; ,
\label{II7} \\
&& \lim_{r\to a^-} K_{33}=a\sin^2 \theta \neq
\lim_{r\to a^+} K_{33}=
\frac{a}{3}\, \sin^2 \theta \; ,  \label{II8}
\end{eqnarray}
where the BH relation $a=9M/4$ was used.

\section*{Appendix C: Dominant order in $R^{(1)}_{\alpha\beta}$}

The second covariant derivatives appearing in Eq.~(\ref{40bis}) are
\def\theequation{C.\arabic{equation}}\setcounter{equation}{0}
\begin{eqnarray}
& & h_{\mu\nu;\alpha\beta}=h_{\mu\nu,\alpha\beta}-
\Gamma^{\sigma}_{\alpha\beta}h_{\mu\nu,\sigma}-
\Gamma^{\sigma}_{\beta\mu}h_{\sigma\nu,\alpha}-
\Gamma^{\sigma}_{\beta\nu}h_{\sigma\mu,\alpha}-
\Gamma^{\sigma}_{\alpha\mu,\beta}h_{\sigma\nu}-
\Gamma^{\sigma}_{\alpha\mu}h_{\sigma\nu,\beta}  \nonumber \\
& & +\Gamma^{\sigma}_{\alpha\beta}\Gamma^{\rho}
_{\sigma\mu}h_{\rho\nu}+
\Gamma^{\sigma}_{\beta\mu}\Gamma^{\rho}
_{\alpha\sigma}h_{\rho\nu}+
\Gamma^{\sigma}_{\beta\nu}\Gamma^{\rho}_{\alpha\mu}h_{\rho\sigma}-
\Gamma^{\sigma}_{\alpha\nu,\beta}h_{\sigma\mu}    \nonumber \\
& & -\Gamma^{\sigma}_{\alpha\nu}h_{\sigma\mu,\beta}+
\Gamma^{\sigma}_{\alpha\beta}\Gamma^{\rho}_{\sigma\nu}h_{\rho\mu}+
\Gamma^{\sigma}_{\beta\nu}\Gamma^{\rho}_{\alpha\sigma}h_{\rho\mu}+
\Gamma^{\sigma}_{\beta\mu}\Gamma^{\rho}_{\alpha\nu}h_{\rho\sigma}
\; .               \label{cov2}                   \end{eqnarray}
Symbolically, we express the various quantities in the last equation
as follows:
\begin{eqnarray}
\label{aux1}
&& \Gamma=\gamma \, \partial \gamma=\mbox{O}(1)      \; , \\
\label{aux2}
&& \gamma \, \partial h=\mbox{O}(1) \; ,   \\
\label{aux3}
&& ( \partial \gamma) h=\mbox{O}( \epsilon)      \; ,   \\
\label{aux4}
&& h \, \partial h=\mbox{O}( \epsilon)   \; ,                 \\
\label{aux5}
&& \Gamma \, \partial h =\mbox{O}(1)   \; ,                \\
\label{aux6}
&& ( \partial \Gamma)h =\mbox{O}( \epsilon)     \;   ,      \\
\label{aux7}
&& \Gamma \, \Gamma \, h=\mbox{O}( \epsilon)      \; .
\end{eqnarray}
By using Eqs.~(\ref{aux1})--(\ref{aux7}) in (\ref{cov2}) and then, in
conjunction with Eq.~(\ref{40bis}), Eq.~(\ref{eq4.1}) follows. The
quantity $\left( h_{\rho\tau,\alpha\beta} +
h_{\alpha\beta,\rho\tau}-h_{\tau\alpha,\beta\rho} 
-h_{\tau\beta,\alpha\rho}\right) $ in Eq.~(\ref{eq4.1}) contains
terms of order O($1/\epsilon$) as well as terms of order O(1).
We retain only the former ones in the linearized Einstein equations to
order O($1/\epsilon$).

\section*{Appendix D: Angle average of $T_\mu^\nu$ in the high
frequency limit}
\def\theequation{D.\arabic{equation}}\setcounter{equation}{0}

Equations (\ref{eq4.15b})--(\ref{eq4.15a}) includes integrating over
the angle $\varphi$ and dividing by the solid angle $4\pi$, thus all
that is left is evaluating the $\theta$ integrals 
(see Ref.~\cite{Wheeler55}). The $\theta$ dependence of $T_\mu^\nu$ comes in 
three forms
\be
\sin^{-2}\theta \left(\Theta^l(\theta)\right)^2,\ \sin^{-2}\theta
\left(\Theta^l(\theta)_{,2}\right)^2 \mbox{ and }
\sin^{-2}\theta\; \Theta^l(\theta) \Theta^l(\theta)_{,22} \; . 
\ee
where
\be
\Theta^l(\theta) = C^{l0} B^l(\theta)
\ee
and
\be
B^l(\theta) \equiv \sin\theta \, \frac{d}{d\theta}\, P^l(\cos\theta) \; .
\ee
The exact integrals are evaluated below with the last equality being
the value used for the high frequency approximation:
\be
\int^\pi_0 \sin^{-2}\theta \left(B^l(\theta)\right)^2 \sin\theta
d\theta = \frac{2 l (l + 1)}{2 l + 1} \approx l \; ,
\ee
\be
\int^\pi_0 \sin^{-2}\theta  
\left(B^l(\theta)_{,\theta}\right)^2 \sin\theta
d\theta = \frac{2 l^2 (l + 1)^2}{2 l + 1} \approx l^3   \; ,
\ee
\be
\int^\pi_0 \sin^{-2}\theta 
B^l(\theta) B^l(\theta)_{,\theta\theta}  \sin\theta
d\theta = -\frac{2 l^3 (l + 1)}{2 l + 1} \approx -\, l^3  \; .
\ee
The normalization constant for $\Theta^l(\theta)$ is found by
requiring
\be
\int^{2\pi}_0 \int^\pi_0  
\left|\Theta^l(\theta)\right|^2    \sin\theta   \, 
d\theta d\varphi = 1   \; .
\ee
Therefore
\be
\left[C^{l0} \right]^2  = \frac{1}{2\pi}\left[ 
 \int^\pi_0  \left(B^l(\theta)\right)^2 \sin\theta
d\theta \right]^{-1} = \frac{1}{2\pi}\left[\frac{4 l^2(l+1)^2}{(2
l-1)(2l+1)(2l+3)}  \right]^{-1}.
\ee
Thus the normalization constant is
\be
C^{l0} = \left[\frac{(2l-1)(2l+1)(2l+3)}{8\pi l^2(l+1)^2} \right]^{1/2}
\approx  \frac{1}{\sqrt{\pi l}} \; .
\ee

\clearpage
\noindent
Fig.~1 Results of the numerical integration for the gravitational geon
differential equations (Eqs.~(\ref{ggwave})--(\ref{ggj})). The
integration was performed from $x = -4$ to $x=35$ with initial
condition $\phi(-4) \equiv \phi_0 = 1.0\times 10^{-4}$. The active
region denoted by the region where the factor $j(x)\, k(x)$ is
positive extends from approximately $x=0$ to $\infty$. Consequently
$\phi(x)$ oscillates out to $\infty$ and the mass of the gravitational
geon (proportional to $(1-k^2)$) is negative for all $x$. Other values
of $\phi_0$ do not qualitatively change the behaviour of the functions
$\phi(x),\, j(x)$ or $k(x)$. 

\vspace{22pt}
\noindent
Fig.~2 Results of the numerical integration for the electromagnetic
geon differential equations (Eqs.~(\ref{eq4.44})--(\ref{eq4.46})). The
initial value of $\phi(-4)$ was $\phi_0 = 9.790419490\times
10^{-5}$. The integration started at $x = -4$ and could not proceed
beyond approximately $x=11$.  The active region began at approximately 
$x=0.1$ and
ended at $x=0.2$. Note that $\phi(x)$ exhibits singular behaviour at
approximately $x=11$.

\vspace{22pt}
\noindent
Fig.~3 Results of the numerical integration for an electromagnetic
geon with an initial value of $\phi_0 = 9.790419489\times 10^{-5}$. For
this case, $\phi(x)$ approaches $-\infty$ at approximately $x=11$.


\clearpage
{\small \begin{thebibliography}{99}

\bibitem{Wheeler55} J. A. Wheeler, {\em Phys. Rev.} {\bf 97} 511 (1955).

\bibitem{Sorkinetal} R. D. Sorkin, R. M. Wald and Z. Z. Jiu, {\em Gen.
                     Rel. Grav.} {\bf 13} 1127 (1981).

\bibitem{Sokolov} S. N. Sokolov, {\em Gen. Rel. Grav.} {\bf 24} 519 (1992).

\bibitem{PowerWheeler} E. A. Power and J. A. Wheeler, {\em Rev. Mod. Phys.}
                       {\bf 29} 480 (1957).

\bibitem{BrillWheeler} R. D. Brill and J. A. Wheeler, {\em Rev. Mod. Phys.}
                       {\bf 29} 465 (1957).

\bibitem{Ernst} F. J. Ernst, {\em Phys. Rev.} {\bf 105} 1662; 1665;
{\em Rev. Mod. Phys.} {\bf 29} 496 (1957).

\bibitem{Wheeler61} J. A. Wheeler, {\em Rev. Mod. Phys.} {\bf 33} 63 (1961).

\bibitem{Wheeler62} J. A. Wheeler, {\em Geometrodynamics} (Academic Press,
                    New York, 1962).

\bibitem{Brill66} D. R. Brill, in {\em Perspectives in Geometry and
Relativity, Essays in honor of V\'{a}clav Hlavat\'{y}}, ed. B. Hoffmann
(Indiana Univ. Press, Bloomington, 1966) p.~38.

\bibitem{ReggeWheeler} T. Regge and J. A. Wheeler, {\em Phys. Rev. D}
{\bf 108} 1063 (1957).

\bibitem{BrillHartle} R. D. Brill and J. B. Hartle, {\em Phys. Rev.} {\bf 135}
                      B271 (1964).

\bibitem{Gerlach} U. H. Gerlach, {\em Phys. Rev. Lett.} {\bf 25} 1771 (1970).

\bibitem{CohenWald} J. M. Cohen and R. M. Wald, {\em J. Math. Phys.}
                    {\bf 13} 543 (1972).

\bibitem{ColemanSmarr} S. Coleman and L. Smarr, {\em Comm. Math. Phys.}
                       {\bf 56} 1 (1977).

\bibitem{geonletter} F. I. Cooperstock, V. Faraoni and G. P. Perry, {\em Mod.
Phys. Lett. A} {\bf 10}, 359 (1995).

\bibitem{Manasse} F. K. Manasse, {\em J. Math. Phys.} {\bf 4} 746 (1963).

\bibitem{Zerilli} F. J. Zerilli, {\em J. Math. Phys.} {\bf 11} 2203 (1970).

\bibitem{Sandberg} V. D. Sandberg, {\em J. Math. Phys.} {\bf 19} 2441 (1978).

\bibitem{Thorne} K. S. Thorne, {\em Rev. Mod. Phys.} {\bf 52} 299 (1980).

\bibitem{Chandrasekhar} S. Chandrasekhar, {\em The Mathematical Theory
of Black Holes} (Clarendon Press, Oxford, 1983).

\bibitem{Isaacson} R. A. Isaacson, {\em Phys. Rev.} {\bf 166}, 1263;
1272 (1968).

\bibitem{BinneyTremaine} J. Binney and S. Tremaine, {\em Galactic
                         Dynamics} (Princeton Univ. Press, Princeton, 1987).

\bibitem{MacCallumTaub} M. A. H. MacCallum and A. H. Taub, {\em Comm. Math.
Phys.} {\bf 30}, 153 (1973).


\bibitem{MTW} C. W. Misner, K. S. Thorne and J. A. Wheeler, {\em
              Gravitation} (Freeman, S. Francisco, 1973).

\bibitem{LandauLifschitz} L. D. Landau and E. M. Lifschitz, {\em
                          The Classical Theory of Fields}, 4th edition
                          (Pergamon Press, Oxford, 1975).

\bibitem{EdelsteinVishveshwara} L. A. Edelstein and C. V. Vishveshwara, 
{\em Phys. Rev.~D} {\bf 1}, 3514 (1970).

\bibitem{Araujo} M. E. de Araujo, {\em Gen. Rel. Grav.} {\bf 18}, 219 (1986);
{\em Gen. Rel. Grav.} {\bf 21}, 323 (1989).

\bibitem{junction} W. B. Bonnor and P. A. Vickers {\em Gen. Rel. Grav.} {\bf
                 13} 29 (1981); G. Darmois {\em M\'emorial des Sciences
                 Mathematiques} (Gauthier--Villars, Paris, 1927),
                 Fasc.~25; S. O'Brien and J. L. Synge {\em Jump Conditions
                 at Discontinuities in General Relativity} (Dublin
                 Institute for Advanced Studies, 1952); A. Lichnerowicz
                 {\em Th\'eories Relativistes de la Gravitation et de
                 l'\'Electromagn\'etisme} (Masson, Paris, 1955).

\bibitem{Israel} W. Israel Nuovo Cimento~B {\bf 44} 1 (1966).


\bibitem{Zerilli2} F. J. Zerilli, {\em Phys. Rev. Lett.} {\bf 24} 737 (1970).

\bibitem{Arnold} V. I. Arnold, {\em Mathematical Methods of Classical
Mechanics} (Springer--Verlag, New York, 1978).

\bibitem{Weinberg} S. Weinberg, {\em Gravitation and Cosmology,
Principles and Applications of the General Theory of Relativity} (John
Wiley \& Sons, New York, 1972).

\bibitem{HTWW} B. K. Harrison, K. S. Thorne, M. Wakano and J. A. Wheeler, 
{\em Gravitation Theory and Gravitational Collapse} (The University of Chicago
Press, 1965), p.~56.

\bibitem{Gibbons} G. W. Gibbons and J. M. Stewart, in {\em Classical
General Relativity}, eds.\ W. B. Bonnor, J. Islam and
M. A. H. MacCallum (Cambridge University Press, 1984) p.~77.

\bibitem{Cooperstock} F. I. Cooperstock, {\em Found. Phys.} {\bf 22}
1011 (1992).

\bibitem{Cooperstock2} F. I. Cooperstock, in {\em Topics in Quantum
                       Gravity and Beyond}, ed F. Mansouri and J. J. Scanio
                       (World Scientific, Singapore, 1993) p.~201.

\bibitem{GursesGursey} M. G\"{u}rses and F. G\"{u}rsey, {\em J. Math. Phys.}
                       {\bf 16} 2385 (1975).

\bibitem{KSMH} D. Kramer, H. Stephani, M. MacCallum and E. Herlt, {\em
               Exact Solutions of Einstein's Field Equations}
               (Cambridge Univ. Press, Cambridge, 1980).
\end{thebibliography}}
\end{document}